\documentclass{article}
\usepackage[dvips]{graphicx}
\usepackage[dvips]{color}
\usepackage{amssymb}
\date{\today}

\makeatletter
\@addtoreset{equation}{section}
\makeatother

\begin{document}
\title{
Large Non-perturbative Effects of Small
$\Delta m^2_{21}/\Delta m^2_{31}$ and $\sin \theta_{13}$ \\
on Neutrino Oscillation and CP Violation in Matter
}
\author{
\sc
{Akira Takamura$^{1,2}$}
\sc
{and}
\sc
{Keiichi Kimura$^2$}
\\
\\
{\small \it $^1$Department of Mathematics,
Toyota National College of Technology}\\
{\small \it Eisei-cho 2-1, Toyota-shi, 471-8525, Japan} \\
{\small \it $^2$Department of Physics, Nagoya University,}
{\small \it Nagoya, 464-8602, Japan}\\
}
\date{}

\maketitle
\begin{abstract}
In the framework of three generations, we consider the CP violation
in neutrino oscillation with matter effects.
At first, we show that the non-perturbative effects of two small
parameters, $\Delta m_{21}^2/\Delta m_{31}^2$ and $\sin \theta_{13}$,
become more than $50\%$ in certain ranges of energy and baseline length.
This means that the non-perturbative effects should be considered
in detailed analysis in the long baseline experiments.
Next, we propose a method to include these effects in approximate
formulas for oscillation probabilities. 
Assuming the two natural conditions, $\theta_{23}=45^\circ$ and 
the fact that the matter density is symmetric, 
a set of approximate formulas, which involve the non-perturbative 
effects, has been derived in all channels.
\end{abstract}

\section{Introduction}

\hspace*{\parindent}
Last year, the direct evidence for neutrino oscillation \cite{MNS}
was found in three kinds of experiments, namely atmospheric
neutrino experiment
\cite{distortion-SuperK},
reactor neutrino experiment \cite{distortion-KamLAND} and
K2K experiment \cite{distortion-K2K}.
In these experiments, the dip of neutrino oscillation and the 
energy dependence of the probability, were observed.
The possibilities of neutrino decay \cite{Neutrino-decay} and
neutrino decoherence \cite{Neutrino-decoherence} are excluded
by these results and it is found that the only solution for
the solar and the atmospheric neutrino problem is neutrino
oscillation.
The observation of the dip also means that
the neutrino experiments herald in a new era of precise
measurements,
because the effect, which disappears by averaging 
out the time-varying part on the neutrino energy, has been observed 
in these experiments for the first time \cite{distortion-SuperK, 
distortion-KamLAND, distortion-K2K}.
Solar neutrino parameters have been also 
accurately determined by recent neutrino experiments such as 
SNO and KamLAND \cite{SNO,KamLAND}.

From the results of the past experiments, it was found that 
the solar neutrino deficit is explained by the Large Mixing 
Angle (LMA) MSW \cite{MSW} solution
\cite{SNO,KamLAND,solar,Bandyopadhyay05},
\begin{eqnarray}
\Delta m_{21}^2 \sim 8.0 \times 10^{-5} {\rm eV^2},
\qquad \sin^2 2\theta_{12} \simeq 0.8,
\end{eqnarray}
where the mass squared difference is defined by
$\Delta m_{ij}^2=m_i^2-m_j^2$.
It was obtained that
\begin{eqnarray}
|\Delta m_{31}^2| \sim 2.0 \times 10^{-3} {\rm eV^2},
\qquad \sin^2 2\theta_{23} \simeq 1.0
\end{eqnarray}
from the atmospheric neutrino experiment \cite{atm}.
Furthermore, the upper bound of the 1-3 mixing angle,
$\sin \theta_{13}$ is given by
\begin{eqnarray}
\sin^2 2\theta_{13} \leq 0.2
\end{eqnarray}
from the reactor experiment \cite{CHOOZ}.
The next step for neutrino physics is the determination of
$\sin\theta_{13}$, the sign of $\Delta m_{31}^2$ and
CP phase $\delta$.
In particular, the measurement of the leptonic CP phase
is one of the most important themes from the viewpoint of
the origin of the universe.
CP violation has been investigated also in quark sector for the 
first time and the Kobayashi-Maskawa theory has been 
established \cite{B-factory}.
However, it has been found that the CP violation in quark sector 
is too small to generate the sufficient baryon number 
in the universe \cite{Cohen93}, because the electroweak 
symmetry breaking is not the first 
phase transition as the Higgs particle is too heavy.
This means that the origin of baryon asymmetry of the universe
is not a CP violation from the KM phase and an extra source 
of CP violation is needed.
One of the alternatives is the generation of a baryon number 
due to the leptonic CP violation \cite{Fukugita86}.
The possibility of this scenario has been investigated 
by many researchers \cite{Luty92}.

In order to attain the next step, the long baseline experiments
like superbeam experiments \cite{Superbeam} and
neutrino factory experiments \cite{NeutrinoFactory} are planned.
In these experiments, the earth matter effects disturb the
observation of the CP violation because the matter in the earth
is not CP invariant and generate the effects of fake CP violation.
Therefore, it is very important to understand the earth matter
effects in neutrino oscillation experiments.

Here, summarizing the results of the atmospheric, 
solar and reactor neutrino experiments,
there are two small parameters 
\begin{eqnarray}
\alpha = \frac{\Delta m_{21}^2}{\Delta m_{31}^2} &\sim& 0.04, \\
s_{13} = \sin\theta_{13} &\leq& 0.23.
\end{eqnarray}
The magnitude of these small parameters is most 
important for measuring the CP violation, because it cannot 
be observed, if one of these parameters vanishes.
As the LMA MSW solution was chosen to explain the results of the solar 
neutrino experiments, $\alpha$ reduced to the largest value
compared to other solutions.
This means that the LMA MSW solution opens the door 
for measuring the leptonic CP violation.
If $s_{13}$ is too small, it will be impossible to 
observe the CP violation.
Therefore the magnitude of $s_{13}$ controls whether 
the leptonic CP violation can be observed or not.

Let us briefly review the approximate formulas using the small
parameter $\alpha$ or $s_{13}$ and the related papers.
At first using the perturbation of oscillation probability 
in $\alpha$, the magnitude of the fake CP violation 
by the matter effects has been investigated 
in \cite{Arafune97,Yasuda99,Koike00,Sato00,Minakata00}.
Furthermore, by expanding the matter potential 
to the Fourier mode, it has been shown in \cite{Koike99,Ota01} 
that the mode with large wavelength mainly contributes 
to the oscillation probability.
Higher order perturbative calculations have been performed by
\cite{Miura01-1,Miura01-2}.
The perturbation in $s_{13}$ has been investigated 
in \cite{Akhmedov01}.
The perturbation in both $\alpha$ and $s_{13}$ has also been studied
in \cite{Cervera00,Freund01} and 
this method has been extended to all channels in \cite{Akhmedov04}.

Next let us review the remarkable features related to the 
leptonic CP violation.
In the case of constant matter density, the notable 
identity $\tilde{J}\tilde{\Delta}_{12}\tilde{\Delta}_{23}
\tilde{\Delta}_{31}=J\Delta_{12}\Delta_{23}\Delta_{31}$ 
has been found in \cite{Naumov92,Harrison00,Parke01},
where $J$ is the Jarlskog factor related to the leptonic CP violation, 
$\Delta_{ij}$ means $\Delta m^2_{ij}/(2E)$ and tilde stands for 
the quantities in matter.
In addition, it has been pointed out that the oscillation 
probability in matter almost coincide with that in vacuum 
under the certain condition, which is called vacuum mimicking 
phenomena, and the method to solve the problem on the 
fake CP violation by using the phenomena is discussed in detail 
\cite{Minakata01,Yasuda01,Lipari01}.
Furthermore, it can be applied to the future long baseline
experiments by using the statistical method explained by 
\cite{Pinney01,Freund01-2}.

In a previous series of papers 
\cite{Yoko02,Kimura02, Yoko00,Takamura04,Kimura04} 
we have considered the three 
generation neutrino oscillation in matter
and have shown that the CP dependence of the oscillation
probabilities are derived exactly \cite{Yoko02}.
In the case that $\nu_e$ is included in the initial or final state,
the CP dependence is given by
\begin{eqnarray}
P(\nu_e \to \nu_e) &=& C_{ee}, \\
P(\nu_\alpha \to \nu_\beta) &=&
A_{\alpha\beta} \cos\delta
+ B_{\alpha\beta} \sin\delta
+ C_{\alpha\beta},
\end{eqnarray}
and in the case that both the initial and final state are
$\nu_\alpha, \nu_\beta=\nu_\mu, \nu_\tau$,
the CP dependence is given by
\begin{eqnarray}
P(\nu_\alpha \to \nu_\beta) &=&
A_{\alpha\beta} \cos\delta
+ B_{\alpha\beta} \sin\delta
+ C_{\alpha\beta}
+ D_{\alpha\beta} \cos2\delta
+ E_{\alpha\beta} \sin2\delta,
\end{eqnarray}
where the coefficients $A_{\alpha\beta} \sim E_{\alpha\beta}$
are independent of the CP phase.
We have also shown that these coefficients can be calculated
exactly in constant matter and then the approximate formulas
are derived in a simple way \cite{Kimura02, Yoko00}.
Furthermore, we proposed a new method for approximating
these coefficients in the case of non-constant matter density
\cite{Takamura04}, and then applied it to the earth matter
\cite{Kimura04}.

In this paper, at first within the framework of two generations, 
it has been shown that perturbation of the small
mixing angle is not effective near the MSW resonance point.
This means that the non-perturbative effects by the small mixing
angle is important in the MSW resonance region.
Next, we consider non-perturbative effects of
$\Delta m_{21}^2/\Delta m_{31}^2$ and $\sin \theta_{13}$
in the three generation neutrino oscillation.
The importance of the non-perturbative effects 
is shown by comparing the exact numerical calculation with
the perturbative expansion of the small parameters.
Furthermore, we consider the method for deriving
the approximate formulas in which the non-perturbative effects
are taken into account.
In our previous paper \cite{Takamura04},
the approximate formulas for $P(\nu_e \to \nu_\mu)$ have been derived.
These formulas are effective for both MSW resonance regions.
However, there is a problem because this method cannot be
extended to other channels $P(\nu_\mu \to \nu_\tau)$ and so on.
In order to solve this problem, we assume the two natural
conditions, $\theta_{23}=45^\circ$ and the symmetric matter
potential.
Under these conditions, we derive the approximate formulas for
all channels, including non-perturbative effects of 
the two small parameters.
These formulas are useful to solve the problem of
parameter degeneracy.-

\section{Non-perturbative Effect by Small Mixing Angle}

\hspace*{\parindent}
In this section, we discuss the perturbative expansion of
a small mixing angle in two generation neutrino oscillation.
Although we discussed the perturbation of small parameters 
in our previous papers \cite{Takamura04, Kimura04}, 
in order to clarify the physical meaning, 
we consider the perturbation due to a small mixing
angle within the framework of two generations.
Then, we show that the perturbation breaks down in the MSW
resonance region even if the mixing angle is small.

\subsection{MSW Resonance of Probability in Two Generations}

\hspace*{\parindent}
In this subsection, we consider the two generation neutrino oscillation
and we choose the energy region and the baseline length in which
the MSW resonance occurs.
Let us start from the Hamiltonian in constant matter
\begin{eqnarray}
H &=& O {\rm diag}(0,\Delta) O^T + {\rm diag}(a,0)
\label{2-generation} \\
 &=& \tilde{O} {\rm diag}(\lambda_1,\lambda_2) \tilde{O}^T
 \label{diagonal},
\end{eqnarray}
where the matter potential is defined by $a=\sqrt{2}G_F N_e $.
$G_F$ is the Fermi constant and $N_e$ is the electron density
in matter.
The matrix $O$ is mixing matrix as
\begin{eqnarray}
O =
\left(
\begin{array}{cc}
\cos\theta & \sin\theta \\
-\sin\theta & \cos\theta
\end{array}
\right),
\end{eqnarray}
where $\Delta=\Delta m^2/2E$ and the quantities with tilde
stand for the quantities in matter.
Diagonalizing (\ref{2-generation}) to (\ref{diagonal}),
the effective masses $\lambda_i (i=1,2)$ and effective mixing angle
$\tilde{\theta}$ are determined.
If we use the notation
$\tilde{\Delta} = \lambda_2-\lambda_1$ as the mass squared difference,
there is a relation between the mass squared difference and
the mixing angles as
\begin{eqnarray}
\frac{\tilde{\Delta}}{\Delta}
= \frac{\sin 2\theta}{\sin 2\tilde{\theta}}
= \sqrt{\left(
\cos 2\theta-\frac{a}{\Delta}
\right)^2 + \sin^2 2\theta}.
\end{eqnarray}
Using these quantities in matter, the oscillation probability is
given by
\begin{eqnarray}
P = \sin^2 2\tilde{\theta} \sin^2 \label{probability}
\frac{\tilde{\Delta}L}{2}.
\end{eqnarray}
The oscillating part with $L/E$ of this probability
becomes large if the condition
\begin{eqnarray}
\sin \frac{\tilde{\Delta} L}{2} \sim 1
\end{eqnarray}
is satisfied.
On the other hand, the condition for the maximal
effective mixing angle is given by
\begin{eqnarray}
\sin 2\tilde{\theta} \sim 1.
\end{eqnarray}
In the case of small mixing angle, this condition 
is rewritten as $a=\Delta \cos 2\theta \sim \Delta$,
and furthermore we define the resonance energy by
\begin{eqnarray}
E \sim \frac{\Delta m^2}{2a}. \label{r-energy}
\end{eqnarray}
We also define the resonance length by
\begin{eqnarray}
L \sim \frac{1}{a\sin\theta} \label{r-length}.
\end{eqnarray}
For the case of 
$\sin \theta = 0.16$, which is the upper bound in the 
CHOOZ experiment, the resonance length is roughly estimated 
as 10000 km.
This means that in near future it is impossible to realize 
the long baseline experiments such that the baseline length
from beam source to the detector is nearly equal to the 
resonance length.
However, it has been shown \cite{Lipari01} that matter 
effects exist even if the baseline length is shorter 
than the resonance length.
Therefore, we use $L=6000$ km as the baseline length 
in the later sections.

\subsection{Perturbation due to Small Mixing Angle}

\hspace*{\parindent}
Next, let us consider the expansion of the effective mass 
$\tilde{\Delta}$ and the effective mixing angle $\sin 2\tilde{\theta}$
by a small mixing angle $\sin \theta$.
We show that
although the effective mass and the effective mixing angle
diverge in the MSW resonance energy region, the oscillation probability, 
which is a function of these two quantities, converges.

At first, the effective mass is expanded as
\begin{eqnarray}
\tilde{\Delta}
= |\Delta-a| + \frac{2a\Delta}{|\Delta-a|}\sin^2 \theta
+ \frac{a^2\Delta^2}{2|\Delta-a|^3}\sin^4 \theta
+ \cdots \label{expansion-mass}.
\end{eqnarray}
One can see from this result that other terms than  
the first term diverge.
The higher order term have larger divergence near the MSW resonance.
The effective mixing angle is expanded as
\begin{eqnarray}
\sin 2\tilde{\theta}
= \frac{\Delta\sin 2\theta}{|\Delta-a|}
\left(
1 - \frac{2a\Delta}{(\Delta-a)^2}\sin^2 \theta
+ \frac{3a^2\Delta^2}{2(\Delta-a)^4}\sin^4 \theta
+ \cdots 
\right). \label{expansion-mixing}
\end{eqnarray}
The condition
\begin{eqnarray}
\sin \theta < \frac{|\Delta-a|}{2\sqrt{a\Delta}}
\end{eqnarray}
is needed for $\sin 2\tilde{\theta}$ to converge the finite value.
However, this condition cannot be satisfied in the MSW resonance region 
defined by $\Delta\sim a$, even if $\sin \theta$ is small.
This means that the above perturbation series diverges.
In the expansion for the effective mass and the
effective mixing angle, the coefficients become large, 
even if these quantities are expanded by the small mixing angle.

Next, let us consider the oscillation probability and
let us demonstrate that the oscillation probability reaches 
a finite value, 
where the divergences due to the effective mass
and the effective mixing angle are canceled out by each other.
Substituting (\ref{expansion-mass}) and (\ref{expansion-mixing})
into (\ref{probability}), we obtain
\begin{eqnarray}
P
&\sim& \frac{\Delta^2\sin^2 2\theta}{(\Delta -a)^2}
\sin^2 \frac{(\Delta-a)L}{2} \nonumber \\
&+&
\frac{\Delta^2\sin^2 2\theta}{(\Delta -a)^2}
\left[
- \frac{4a\Delta\sin^2\theta}
{(\Delta-a)^2}\sin^2 \frac{(\Delta-a)L}{2}
+ \frac{a\Delta L\sin^2\theta}{\Delta-a}\sin (\Delta-a)L
\right] + \cdots.
\end{eqnarray}
In the limit, $\Delta \sim a$, it is found that
the oscillation probability becomes finite as
\begin{eqnarray}
P
&\sim& \cos^2\theta
\left(
\sin^2\theta a^2 L^2 -\frac{1}{3}\sin^4\theta a^4L^4 +\cdots
\right) \label{expansion-resonance}.
\end{eqnarray}
From this equation, the oscillation probability becomes finite
and the perturbation is a good approximation
if

\begin{eqnarray}
L < \frac{1}{a\sin\theta} \label{condition}.
\end{eqnarray}
As you see from (\ref{r-length}),
this is the condition that the baseline length
is shorter than the resonance length.

Next, let us investigate the magnitude of non-perturbative effects
numerically.
We use the following parameters,
$\Delta m^2=2.0 \times 10^{-3}~{\rm eV^2}$ and
$\sin \theta =0.16$.
We set the baseline length, $L=6000~{\rm km}$ and
the energy region, $1~{\rm GeV} \leq E \leq 50~{\rm GeV}$,
to include the MSW resonance energy.
Furthermore we choose a density of 
$\rho= 4~{\rm g/cm^3}$.

\begin{figure}[!htb]
\begin{tabular}{cc}
(a) Level crossing & (b) Probability \\
\resizebox{77mm}{!}{\includegraphics{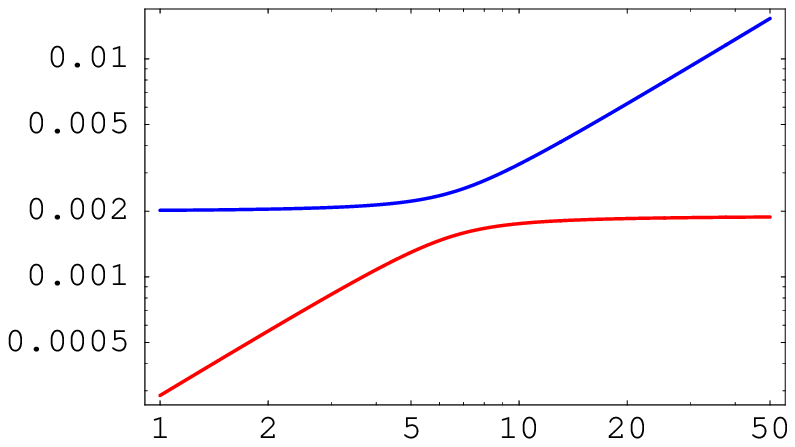}} &
\resizebox{77mm}{!}{\includegraphics{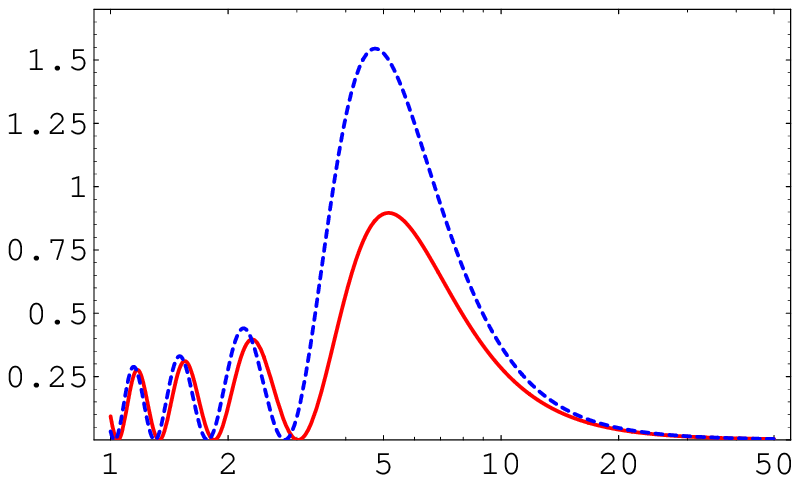}}
\end{tabular}
\caption{
Comparison of the perturbative value
with the exact one in the two generation neutrino oscillation
probability.
In (a), the energy dependence of two eigenvalues is plotted. 
In (b), the dotted and solid line show the values by the perturbative 
and numerical calculations, respectively.}
\end{figure}

At first, in figure 1(a) the level crossing of 
two eigenvalues is plotted.
It is shown that the crossing energy is about 6-7 GeV, which 
corresponds to the MSW resonance energy.
Next, in figure 1(b) we compare the oscillation probability calculated 
by perturbation with the one by numerical calculation.
These figures show that the perturbation breaks down
around the MSW resonance energy.
The results of this subsection are summarized as
\begin{enumerate}
\item The perturbative expansion in the small mixing angle 
breaks down around the MSW resonance because the perturbation
because the perturbation series diverges.
The coefficients of this expansion become larger
around the MSW resonance.
The divergence included in the effective mass 
cancels with that in the effective mixing angle, 
and as a result, the value of
the oscillation probability reaches a finite value.
Term of eq. (\ref{expansion-mass}) and (\ref{expansion-mixing}) 
cancel with each other.

\item Although the divergences of the effective mass and
the effective mixing angle in the perturbative expansion 
cancel in the oscillation probability,
the finite value of the probability differs from that 
by numerical calculation.
The perturbation around the MSW resonance energy
becomes a good approximation, if the baseline length 
is shorter than the resonance length as seen from (\ref{condition}).
However, we need to take higher order terms of the perturbation into account, 
when the baseline length is longer,
namely when non-perturbative effects become important.
\end{enumerate}

\section{Extension of method to Approximate
Oscillation Probabilities}

\hspace*{\parindent}
In this section, we consider the matter effects in
three generation neutrino oscillation.
At first, we review that the 2-3 mixing angle
$\theta_{23}$
and the CP phase $\delta$ can be separated from matter
effects in the oscillation probability \cite{Yoko02}.
This means that the matter effects appear through the
remained four parameters.
Furthermore, these four parameters can be separated
to two set of parameters and each set is related to
the phenomena in low and high energy as
\begin{eqnarray}
&& (\theta_{12}, \Delta m^2_{21})
: {\rm Low~Energy~Phenomenon} \\
&& (\theta_{13}, \Delta m^2_{31})
: {\rm High~Energy~Phenomenon}.
\end{eqnarray}
This separation means that the parameters for the
solar neutrino and those for the atmospheric neutrino 
are almost independent to each other.
We propose the method deriving the approximate
formulas simply
by using this feature.

\subsection{Definition of Low and High Energy Regions}

\hspace*{\parindent}
In this subsection, we define the low energy
and the high energy Hamiltonians in the small quantity
limit when $s_{13}$ or $\alpha$ approximate zero.
Although these Hamiltonian have been already
introduced in our earlier papers \cite{Takamura04, Kimura04},
we review them here, as they are used in later section.
\par
It is noted that $H(t)$ satisfies the relation
\begin{equation}
H(t) = O_{23} \Gamma H'(t)\Gamma^\dagger O_{23}^T
\end{equation}
where $H'$ is given by
\begin{equation}
H' = O_{13}O_{12}{\rm
diag}(0,\Delta_{21},\Delta_{31})O_{12}^TO_{13}^T
+ {\rm diag}(a(t),0,0).
\label{primed-Hamiltonian}
\end{equation}
This means that the 1-2 and 1-3 mixing angles are
separated from the 2-3 mixing and the CP phase, 
as explained in detail in Appendix A.
In this Appendix A, we derive the same result as
that derived from this section from another point of
view.
Taking the limit $s_{13} \to 0$, the Hamiltonian
reduces to the two generation Hamiltonian as
\begin{eqnarray}
H^\ell &=& \lim_{s_{13} \to 0} H' \\
&=& O_{12}{\rm diag}(0,\Delta_{21},\Delta_{31})
O_{12}^T
+ {\rm diag}(a(t),0,0) \\
&=&
\left(
\begin{array}{ccc}
\Delta_{21} s_{12}^2 + a(t)
& \Delta_{21}s_{12}c_{12}
& 0 \\
\Delta_{21}s_{12}c_{12}
& \Delta_{21} c_{12}^2
& 0 \\
0
& 0
& \Delta_{31}
\end{array}
\right) \label{low-Hamiltonian}
\end{eqnarray}
This means that the third generation is now
separated from the first and the second generation.
As seen from this Hamiltonian (\ref{low-Hamiltonian}),
the components except for $H_{\tau\tau}^\ell$, depend
only on $(\theta_{12}, \Delta_{21})$.
We call $H^\ell$ the low energy Hamiltonian.
On the other hand, taking the limit $\alpha \to 0$,
the Hamiltonian reduces to the two generation
Hamiltonian as
\begin{eqnarray}
H^h &=& \lim_{\alpha \to 0} H' \\
&=& O_{13}{\rm diag}(0,0,\Delta_{31})O_{13}^T
+ {\rm diag}(a(t),0,0) \\
&=&
\left(
\begin{array}{ccc}
\Delta_{31} s_{13}^2 + a(t)
& 0
& \Delta_{31}s_{13}c_{13} \\
0
& 0
& 0 \\
\Delta_{31}s_{13}c_{13}
& 0
& \Delta_{31} c_{13}^2
\end{array}
\right) \label{high-Hamiltonian}.
\end{eqnarray}
This means that the second generation is also
separated from the two others.
This Hamiltonian (\ref{high-Hamiltonian}) is expressed
by only the parameters $(\theta_{13}, \Delta_{31})$.
We call $H^h$ high energy Hamiltonian.
Next, let us define the high and low energy
regions described by $H^h$ and $H^\ell$.
We first calculate the MSW resonance energy because
the MSW effect is the most important in matter
effects.
In the case of $L=6000~{\rm km}$, which we use later,
the average matter potential is calculated as
$\rho= 3.91~{\rm g/cm^3}$.
By using this value, we obtain the high energy MSW
resonance as $E^h = \Delta m_{31}^2/a \simeq 5~{\rm GeV}$
and the low energy MSW resonance as
$E^\ell = \Delta m_{21}^2/a \simeq 0.2~{\rm GeV}$.
From these results, we regard $E \sim 1 {\rm GeV}$
as the boundary energy of low and high energy regions.
Therefore, we define the high as $E \geq 1~{\rm GeV}$ 
and the low energy regions as $E \leq 1~{\rm GeV}$.

\subsection{Order Counting of Amplitude on $\alpha$
and $s_{13}$}

\hspace*{\parindent}
In this subsection, we investigate how the amplitude $S'$, 
which is defined by the primed Hamiltonian 
(\ref{primed-Hamiltonian}), depends on the two small parameters 
$\alpha$ and $s_{13}$.
Before, we have already clarified some general
features of $S'$ related to the order of $\alpha$ 
and $s_{13}$, and the dependences on $s_{13}$ and 
$\alpha$ for particular amplitudes $S'_{\mu e}$ and 
$S'_{\tau e}$ have been given
in our previous papers \cite{Takamura04,Kimura04}.
We investigate now the dependences on $s_{13}$ and
$\alpha$ for all amplitudes.
\par
At first, we represent the explicit form of the
Hamiltonian, when the 2-3 mixing angle and the CP
phase are factored out as
\begin{eqnarray}
H'(t) &=& O_{13}O_{12}{\rm
diag}(0,\Delta_{21},\Delta_{31})
O_{12}^TO_{13}^T
+ {\rm diag}(a(t),0,0) \\
&=& \left(
\begin{array}{ccc}
\Delta_{21} c_{13}^2s_{12}^2 + \Delta_{31} s_{13}^2 +
a(t)
& \Delta_{21}c_{13}s_{12}c_{12}
& -\Delta_{21}c_{13}s_{13}s_{12}^2 +
\Delta_{31}s_{13}c_{13} \\
\Delta_{21}c_{13}s_{12}c_{12}
& \Delta_{21} c_{12}^2
& -\Delta_{21} s_{13}s_{12}c_{12} \\
-\Delta_{21}c_{13}s_{13}s_{12}^2 +
\Delta_{31}s_{13}c_{13}
& -\Delta_{21} s_{13}s_{12}c_{12}
& \Delta_{21} s_{13}^2s_{12}^2 + \Delta_{31} c_{13}^2
\end{array}
\right).
\end{eqnarray}
The components of this Hamiltonian depend on
$\alpha$ and $s_{13}$ as
\begin{eqnarray}
H'(t) &=& \left(
\begin{array}{ccc}
O(1) & O(\alpha) & O(s_{13}) \\
O(\alpha) & O(\alpha) & O(\alpha s_{13}) \\
O(s_{13}) & O(\alpha s_{13})& O(1)
\end{array}
\right).
\end{eqnarray}
From this result, we can see that non-diagonal
components are
small compared to the diagonal components.
We also understand that $H'_{\mu\tau}$ is the
smallest component
and $H'_{e\mu}, H'_{e\tau}$ are the next smaller
components.
We should note the salient feature that the result of
this order counting holds in $H^n$ for arbitrary $n$.
Namely, we obtain
\begin{eqnarray}
(H'(t))^n &=& \left(
\begin{array}{ccc}
O(1) & O(\alpha) & O(s_{13}) \\
O(\alpha) & O(\alpha^2) & O(\alpha s_{13}) \\
O(s_{13}) & O(\alpha s_{13})& O(1)
\end{array}
\right) \quad {\rm for} \quad n = 1, 2, \cdots.
\end{eqnarray}
According to this result, the order of the
amplitude $S'(t)$
for two small parameters $\alpha$ and $s_{13}$ is
given by
\begin{eqnarray}
S'(t)
= {\rm T}\exp\left\{-i\int H'(t) dt\right\}
= \left(
\begin{array}{ccc}
O(1) & O(\alpha) & O(s_{13}) \\
O(\alpha) & O(1) & O(\alpha s_{13}) \\
O(s_{13}) & O(\alpha s_{13})& O(1)
\end{array}
\right). \label{feature-1}
\end{eqnarray}
This result is almost the same as that of the original
Hamiltonian.
Furthermore, we consider the general features
derived from the original Hamiltonian.
The $\theta_{13}$ dependence of this Hamiltonian is
described as
\begin{equation}
H' =
\left(
\begin{array}{ccc}
{\rm even} & {\rm even} & {\rm odd} \\
{\rm even} & {\rm even} & {\rm odd} \\
{\rm odd} & {\rm odd} & {\rm even}
\end{array}
\right)
\end{equation}
and this dependence does not change for $(H')^n$,
because
\begin{equation}
(H')^n =
\left(
\begin{array}{ccc}
{\rm even} & {\rm even} & {\rm odd} \\
{\rm even} & {\rm even} & {\rm odd} \\
{\rm odd} & {\rm odd} & {\rm even}
\end{array}
\right) \quad {\rm for} \quad n = 1, 2, \cdots.
\end{equation}
Due to this result, the amplitude $S'(t)$ has the
same structure,
\begin{eqnarray}
S'
= {\rm T}\exp\left\{-i\int H'(t) dt\right\}
= \left(
\begin{array}{ccc}
{\rm even} & {\rm even} & {\rm odd} \\
{\rm even} & {\rm even} & {\rm odd} \\
{\rm odd} & {\rm odd} & {\rm even}
\end{array}
\right) \label{feature-2}.
\end{eqnarray}
This is a general feature, which holds in
arbitrary matter profile.

\subsection{Proposal of Simple Method}

\hspace*{\parindent}
In the previous subsection, we have shown
the general features (\ref{feature-1}) and
(\ref{feature-2}) for the amplitude $S'(t)$
related to the almost vanishing parameters
$s_{13}$ and $\alpha$.
However, we cannot calculate $S'(t)$ by using only
this features.
In this subsection, we propose a generalized method 
for the calculation.
Let us consider if there is an approximation
available for both region, low and high energy.
After expanding the amplitude $S'$ on the two small
parameters $\alpha$ and $s_{13}$, we can arrange this as
\begin{eqnarray}
S'
&=& O(1) + O(\alpha) + O(s_{13}) +
O(\alpha^2) + O(\alpha s_{13}) + O(s_{13}^2)
+ \cdots \label{proposal-1} \\
&=& \Big(O(1) + O(\alpha) + O(\alpha^2) + \cdots \Big)
+
\Big(O(1) + O(s_{13}) + O(s_{13}^2) + \cdots \Big)
\nonumber \label{proposal-2} \\
&-& O(1) + O(\alpha s_{13}) + O(\alpha^2 s_{13}) +
O(\alpha s_{13}^2) +
\cdots \label{proposal-3} \\
&=& \lim_{s_{13} \to 0}S'
+ \lim_{\alpha \to 0}S'
- \lim_{\alpha,s_{13} \to 0}S'
+ O(\alpha s_{13}) + O(\alpha^2 s_{13}) + \cdots \\
&=& S^\ell + S^h - S^d + O(\alpha s_{13}) + O(\alpha^2
s_{13})
+ O(\alpha s_{13}^2) + \cdots \label{proposal-4},
\end{eqnarray}
where $S^\ell, S^h$ and $S^d$ are defined by
\begin{eqnarray}
S^\ell
&=& \lim_{s_{13} \to 0}S'
= {\rm T}\exp\left\{-i\int H^\ell dt\right\}
\label{low-S} \\
S^h
&=& \lim_{\alpha \to 0}S'
= {\rm T}\exp\left\{-i\int H^h dt\right\}
\label{high-S} \\
S^d
&=& \lim_{\alpha,s_{13} \to 0}S'
= {\rm diag}\left(\exp\left\{-i\int a(t)dt\right\},1,
e^{-i\Delta_{31}L}\right),
\end{eqnarray}
respectively.
$S^\ell$ ($S^h$) corresponds to the amplitudes, which
gives the
main contribution in low (high) energy.
The term $S^d$ counts twice, because contributions
to this term comes from both, low energy and high
energy terms.
Therefore, we subtract this contribution and
approximate the amplitude as
\begin{eqnarray}
S' \sim S^\ell + S^h - S^d \label{new-proposal}
\end{eqnarray}
ignoring higher order terms.
Let us discuss this approximation, which is used to
derive our main result here.
\par
In (\ref{proposal-1})-(\ref{proposal-4}),
the higher order terms in $\alpha$ and $s_{13}$
are included in $S^\ell$ and $S^h$.
The reason for including the higher order terms is to
take into account non-perturbative effects, which become
important in the low and high energy MSW resonance region as
discussed in section 2.
On the other hand, we ignore those higher order
terms, which are proportional to both $\alpha$ and $s_{13}$.
For example, in the case of second order of the
small parameters, $\alpha$ and $s_{13}$,
we ignore only the mixed $O(\alpha s_{13})$ term among
the three terms with second order
$O(\alpha^2), O(s_{13}^2)$ and $O(\alpha s_{13})$.
This procedure is more appropriate than the usual
perturbation,
because both non-perturbative effects on a small
$\alpha$
in the low energy region and on a small $s_{13}$ in
the high energy region
can be included in our approximation.
However, as the derivation of the approximation
(\ref{new-proposal})
is not exact, we need to check this later numerically.
In the previous subsection, the parity of the matrix
elements
related to $s_{13}$ has been derived.
The equations (\ref{feature-1}), (\ref{feature-2}) and
(\ref{new-proposal})
lead to the magnitude of the correction for
the amplitudes as
\begin{eqnarray}
S'_{\mu e} &=& S^\ell_{\mu e} + O(\alpha s_{13}^2)
\label{S_me} \\
S'_{\tau e} &=& S^h_{\tau e} + O(\alpha s_{13})
\label{S_te} \\
S'_{\tau \mu} &=& O(\alpha s_{13}) \label{S_tm} \\
S'_{ee} &=& S^\ell_{ee} + S^h_{ee} - S^d_{ee} +
O(\alpha s_{13}^2)
\label{S_ee} \\
S'_{\mu\mu} &=& S^\ell_{\mu\mu} + O(\alpha s_{13}^2)
\label{S_mm} \\
S'_{\tau\tau} &=& S^h_{\tau\tau} + O(\alpha s_{13}^2)
\label{S_tt}.
\end{eqnarray}
If we ignore the higher order terms which are
proportional to both,
$\alpha$ and $s_{13}$, in these equations,
we obtain approximate formulas by using the two
generation
amplitudes.
The main term for $S'_{\mu e}, S'_{\mu\mu}$ is
approximated
by the low-energy amplitude as seen from (\ref{S_me})
and
(\ref{S_mm}).
On the other hand, the main terms for $S'_{\tau
e}$ and $S'_{\tau\tau}$ are approximated by the 
high-energy amplitude as derived from (\ref{S_te}) 
and (\ref{S_tt}), and these features
come from eq. (\ref{feature-1}).
As seen from (\ref{S_me})-(\ref{S_tt}),
these are expressed by only two generation amplitudes
and have the advantage of simplicity.
The precision of the approximation depends on
the values of $s_{13}$ and $\alpha$.
If the value of $s_{13}$ is smaller than the upper
bound derived by the CHOOZ experiment, the precision of
approximation becomes better.
It should be mentioned that the method described
in this subsection does not need the assumption of 
constant matter density.
\par
Next, we show that the results using the approximate 
formulas (\ref{S_me})-(\ref{S_tt}) are in excellent 
agreement with the numerical calculations.
We choose the Preliminary Reference Earth Model (PREM)
as an earth matter density model and compare the
amplitudes
in all channels calculated from our approximate
formulas
with the numerical calculation.
Here,
$\Delta m_{21}^2=8.3 \times 10^{-5}~{\rm eV^2},
\Delta m_{31}^2=2.0 \times 10^{-3}~{\rm eV^2},
\sin^2 2\theta_{12}=0.8$ and $\sin \theta_{13}=0.23$
are chosen.
Furthermore, we set the baseline length as $L=6000~{\rm km}$,
a length, for which the MSW effect becomes
significant, and the energy region as
$1~{\rm GeV} \leq E \leq 20~{\rm GeV}$, for which the
MSW resonance energy appears.

\begin{figure}[!htb]
\begin{tabular}{ccc}
$|S'_{\mu e}|$ & $|S'_{\tau e}|$ & $|S'_{\tau\mu}|$ \\
\resizebox{50mm}{!}{\includegraphics{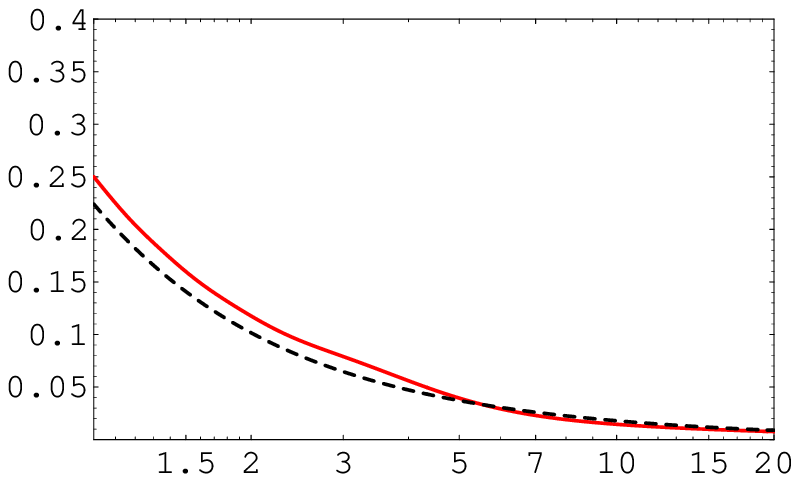}} &
\resizebox{50mm}{!}{\includegraphics{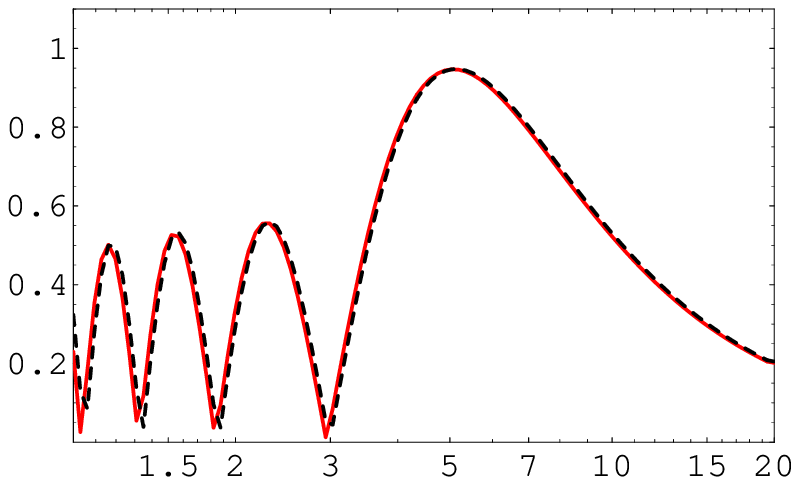}} &
\resizebox{50mm}{!}{\includegraphics{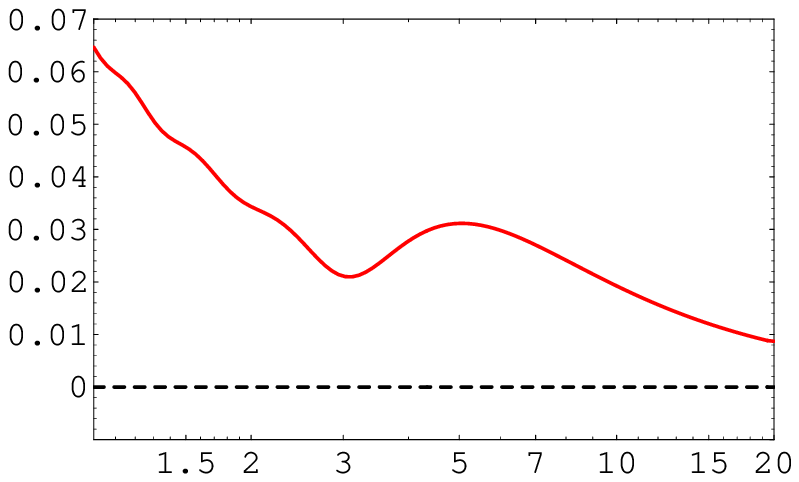}} \\
$|S'_{ee}|$ & $|S'_{\mu\mu}|$ & $|S'_{\tau\tau}|$ \\
\resizebox{50mm}{!}{\includegraphics{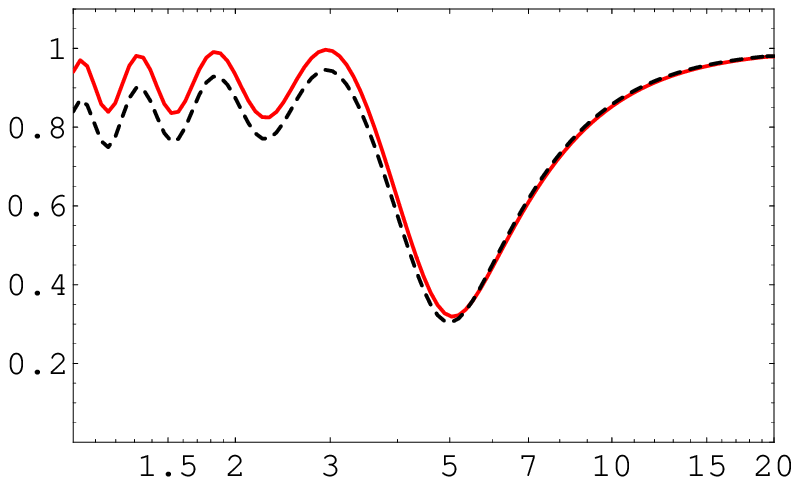}} &
\resizebox{50mm}{!}{\includegraphics{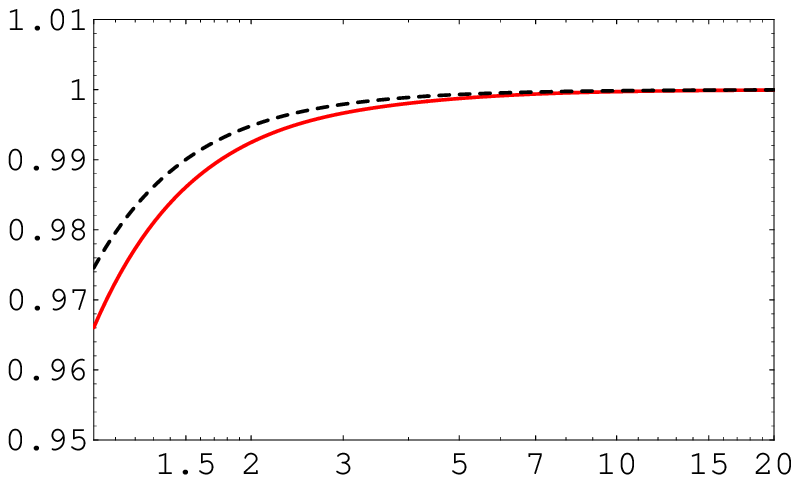}} &
\resizebox{50mm}{!}{\includegraphics{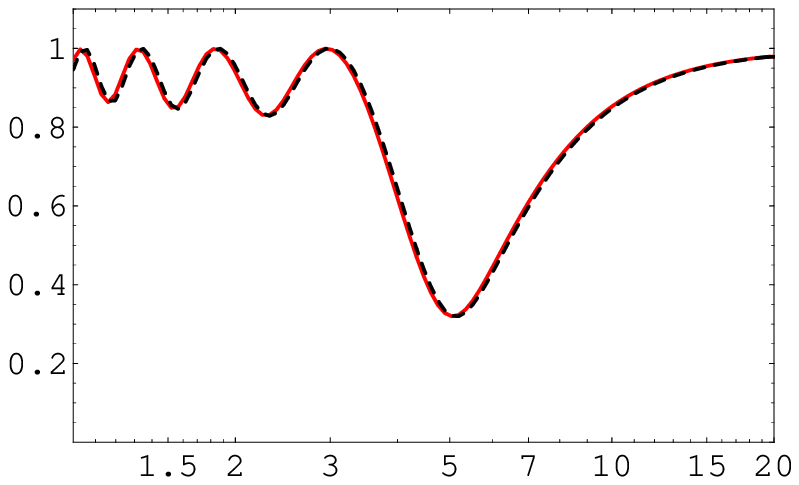}}
\end{tabular}
\caption{
Comparison of our approximate formulas with the
numerical
calculation.
In these figures the absolute value of the
amplitudes in all channels is plotted in order to 
compare our formulas with numerical calculation.
The solid lines show the approximate probabilities
calculated from (\ref{S_me})-(\ref{S_tt}) and the
dashed lines
show the probabilities in the numerical calculation.
}
\end{figure}

We compare our formulas with the numerical calculation
in Figure 2.
One can see in the following that some remarkable
features occur.
At first, the four amplitudes
$|S'_{\mu e}|, |S'_{ee}|, |S'_{\mu\mu}|$ and
$|S'_{\tau\tau}|$
coincide with the numerical calculation with a good
precision.
This happens, because there is no first order
correction of $s_{13}$ from (\ref{S_me}) and
(\ref{S_ee})-(\ref{S_tt}).
Next, the low-energy part of $|S'_{\tau e}|$
differs from the numerical calculation only a little,
which can be understood from the eq. (\ref{S_te}).
Furthermore, our approximation for $|S'_{\tau\mu}|$
is not at all in agreement with the numerical
calculation.
Although the value of this amplitude is exactly zero
in our approximation as seen from (\ref{S_tm}),
the actual magnitude of this amplitude attains $0.02$
in the low energy region from Figure 1.
It is noted that this value is almost the same as
the value expected from the order counting 
$O(\alpha s_{13}) \sim 0.01$.
Next, we would like to derive the approximate formulas
of the oscillation probabilities from the amplitudes
obtained here, however, there is a problem.
As seen from eqs. (\ref{P_mm})-(\ref{E_mt}) in
Appendix A, we cannot obtain the approximate formulas
for the CP dependence of the probabilities
$P(\nu_\mu \to \nu_\mu), P(\nu_\mu \to \nu_\tau)$
and $P(\nu_\tau \to \nu_\tau)$.
The reason is that the CP dependence in these channels
is directly proportional to $S'_{\mu\tau}$.
However there is a method to calculate these indirectly 
by using the unitarity, even if we cannot directly 
obtain the amplitude $S'_{\mu\tau}$, as we will show 
in section 4.

\subsection{Discussion}

\hspace*{\parindent}
In this subsection, let us reconsider the method
proposed in the previous subsection in more detail.
In (\ref{proposal-1})-(\ref{proposal-4}),
we ignored the terms of the order $O(\alpha s_{13})$
for the amplitude $S'$.
The reader probably wonder, why we ignore the
terms of order $O(\alpha s_{13})$
for the amplitude $S'$, but not for other quantities,
like for example $H'$ and $P$.
Let us demonstrate the case of using the physical
quantity $Q$.
Expanding $Q$ on $\alpha$ and $s_{13}$, we obtain
\begin{eqnarray}
Q = Q^\ell + Q^h - Q^d
+ O(\alpha s_{13}) + O(\alpha^2 s_{13}) + \cdots
\end{eqnarray}
by the same procedure as
(\ref{proposal-1})-(\ref{proposal-4}).
If we neglect the higher order terms like $O(\alpha
s_{13})$, we can approximate $Q$ as
\begin{eqnarray}
Q \sim Q^\ell + Q^h - Q^d.
\end{eqnarray}
As in the case of the approximated amplitude
defined in the previous subsection,
$Q^\ell=Q^\ell(\theta_{12},\Delta_{21})$ is the main
term in low-energy and $Q^h=Q^h(\theta_{13},\Delta_{31})$
is the main term in high-energy.
$Q^d$ is the double counting term.
It is a method to be able to take into account 
non-perturbative effects in both of the two MSW 
resonance regions.
In principle, this method is effective whatever we
choose for the quantity $Q$, there is just a difference in
simplicity and the magnitude of error, as discussed
in the following.
\par
We consider the Hamiltonian $H'$ as $Q$. Namely,
$H'$ can be approximated as
\begin{eqnarray}
H' \sim H^\ell + H^h - H^d,
\end{eqnarray}
where the double counting term is given by
\begin{eqnarray}
H^d = {\rm diag}\left(a(t),0,\Delta_{31}\right).
\end{eqnarray}
There is a problem, because approximation became too
simple: The form of the solution for the amplitude is
given by
\begin{eqnarray}
S' \sim {\rm T}\exp\left\{-i\int (H^\ell + H^h - H^d)
dt\right\},
\end{eqnarray}
and we cannot simplify this amplitude without
calculation
of the commutator of $H^\ell$ and $H^h$.
Thus, the direct application of our method for the
Hamiltonian needs other approximations to estimate the
amplitude and this is not effective from the point of the
simplicity.
Especially, the amplitudes cannot be calculated
within the framework of the two generation approximation 
although the precision of this approximation was good.
\par
Next, let us consider the probability $P$ as the
quantity $Q$.
In this case, we can approximate as
\begin{eqnarray}
P \sim P^\ell + P^h - P^d,
\end{eqnarray}
where $P^\ell$ and $P^h$ are given by
\begin{eqnarray}
P^{\ell(h)} =
\left|{\rm T}\exp\left\{-i\int H^{\ell(h)}
dt\right\}\right|^2,
\end{eqnarray}
and $P^d$ is the identity matrix.
As an example, we consider $P(\nu_e \to \nu_\mu)$.
The CP phase $\delta$ dependence is given by
\begin{eqnarray}
P(\nu_e \to \nu_\mu) =
A_{e\mu} \cos\delta
+ B_{e\mu} \sin\delta
+ C_{e\mu},
\end{eqnarray}
where the coefficients $A_{e\mu}$ and $B_{e\mu}$
determine the magnitude of the CP violation.
On the other hand, the CP violation becomes zero in
the limit,
$\alpha \to 0$ or $s_{13} \to 0$, as seen from
\begin{eqnarray}
A_{e\mu} = O(\alpha s_{13}),\quad B_{e\mu} = O(\alpha
s_{13}).
\end{eqnarray}
Namely, we obtain
\begin{eqnarray}
A_{e\mu}^\ell=A_{e\mu}^h=A_{e\mu}^d=0, \quad
B_{e\mu}^\ell=B_{e\mu}^h=B_{e\mu}^d=0
\end{eqnarray}
and therefore we cannot calculate quantities like
the CP violation,
because it is the effects of three generations
in this approximation.
This result holds for all channels.
\par
To summarize this subsection, if we choose the
Hamiltonian $H'$ as $Q$,
the precision of approximation is good, but the
calculation
is not so simple compared to the exact calculation.
If we choose the probability $P$ as $Q$,
we cannot calculate three generation effects like CP
violation.
On the other hand, if we choose the amplitude $S'$ as
$Q$, we can calculate the three generation effects
like CP violation within the framework of two
generation approximation.

\section{Approximate Formulas for Oscillation
Probabilities}

\hspace*{\parindent}
In this section, we calculate  the CP dependent terms 
from $\nu_\mu$ to $\nu_\mu$ and so on, not
determined by the method in the previous section, 
by using the unitarity.
After that, we derive the approximate formulas of the
oscillation probabilities
$P(\nu_\alpha \to \nu_\beta)$ in arbitrary matter
profile without using $S'_{\mu\tau}$ directly.
Namely, we derive the approximate formulas in all
channels
by our new method.

\subsection{Unitarity Relation}

\hspace*{\parindent}
We cannot calculate the amplitude $S'_{\mu\tau}$ in
the method
introduced in the previous section.
The reason is that the amplitude $S'_{\mu\tau}$ is a
very small
quantity, which has an order of $O(\alpha s_{13})$.
As seen from (\ref{P_mm})-(\ref{E_mt}) in Appendix A,
it seems that the approximate formulas, including CP
violation,
of three channels, $P(\nu_\mu \to \nu_\mu), P(\nu_\tau
\to \nu_\tau)$
and $P(\nu_\mu \to \nu_\tau)$ cannot be derived
without directly calculating the amplitude
$S'_{\mu\tau}$.
However in this subsection we show, that we can
derive these probabilities
without directly calculating this amplitude, if we
assume the two natural conditions,
\begin{eqnarray}
s_{23} &\simeq& c_{23}, \label{assumption-1} \\
S'_{\alpha\beta} &\simeq& S'_{\beta\alpha}
\label{assumption-2}.
\end{eqnarray}
The first condition is supported by the best fit
value of atmospheric neutrino experiments \cite{atm} 
and the second condition holds
in one-dimensional models of the earth matter density
like PREM or ak-135f.
Accordingly, the error due to the difference between
these conditions
and the real situations is considered to be relatively
small.
We perform the analysis under these two conditions
in the following.

At first, we obtain
\begin{eqnarray}
B_{\mu\mu}
= - 2 {\rm Im}[({S}'_{\mu \mu} c_{23}^2 + S'_{\tau
\tau} s_{23}^2)^{*}
({S}'_{\tau \mu} - S'_{\mu \tau})] c_{23} s_{23}
= 0 \label{relation-1}
\end{eqnarray}
from (\ref{B_mm}) and (\ref{assumption-2})
in the case of the symmetric matter density.
In the same way, we obtain
\begin{eqnarray}
B_{\tau\tau}
= 2 {\rm Im}[({S}'_{\mu \mu} s_{23}^2 + S'_{\tau \tau}
c_{23}^2)^{*}
({S}'_{\tau \mu} - S'_{\mu \tau})] c_{23} s_{23}
= 0 \label{relation-2}
\end{eqnarray}
from (\ref{B_tt}) and (\ref{assumption-2}).
Furthermore, in the case of the symmetric matter
density
and the maximal 2-3 mixing angle 45$^\circ$,
we also obtain
\begin{eqnarray}
A_{\mu\tau}
= - 2 {\rm Re}[({S}'_{\mu \mu} - S'_{\tau \tau})^{*}
({S}'_{\tau \mu} c_{23}^2 - S'_{\mu \tau}
s_{23}^2)]c_{23} s_{23}
= 0 \label{relation-3}
\end{eqnarray}
from (\ref{A_mt}) and (\ref{assumption-2}).
Let us here consider now, how the oscillation
probabilities are derived,
which are related to the amplitude $S'_{\mu\tau}$ but
have not been determined
in the previous section,.
At first, in the probability,
\begin{eqnarray}
P(\nu_\mu \to \nu_\mu)
= A_{\mu\mu} \cos\delta
+ B_{\mu\mu} \sin\delta
+ C_{\mu\mu}
+ D_{\mu\mu} \cos2\delta
+ E_{\mu\mu} \sin2\delta,
\end{eqnarray}
the coefficient proportional to $\cos\delta$ can be
calculated as
\begin{eqnarray}
A_{\mu\mu}
= - A_{\mu e} - A_{\mu\tau}
\simeq -A_{e\mu}
\simeq -2{\rm Re}[S^{\ell *}_{\mu e} S^h_{\tau
e}]s_{23}c_{23}
\end{eqnarray}
from (\ref{relation-3}) and the unitarity relation.
Next, let us turn to the probability $P(\nu_\tau \to
\nu_\tau)$.
In the probability,
\begin{eqnarray}
P(\nu_\tau \to \nu_\tau)
= A_{\tau\tau} \cos\delta
+ B_{\tau\tau} \sin\delta
+ C_{\tau\tau}
+ D_{\tau\tau} \cos2\delta
+ E_{\tau\tau} \sin2\delta,
\end{eqnarray}
the coefficient of $\cos\delta$ can be calculated as
\begin{eqnarray}
\hspace*{-0.5cm}
A_{\tau\tau}
= -A_{\tau e} -A_{\tau\mu}
\simeq -A_{e\tau}
\simeq 2{\rm Re}[S^{\ell *}_{\mu e} S^h_{\tau
e}]s_{23}c_{23}
\end{eqnarray}
from (\ref{relation-3}) and the unitarity relation.

Finally, let us calculate the probability
$P(\nu_\mu \to \nu_\tau)$.
In the probability,
\begin{eqnarray}
P(\nu_\mu \to \nu_\tau)
= A_{\mu\tau} \cos\delta
+ B_{\mu\tau} \sin\delta
+ C_{\mu\tau}
+ D_{\mu\tau} \cos2\delta
+ E_{\mu\tau} \sin2\delta,
\end{eqnarray}
the coefficient of $\sin\delta$ becomes
\begin{eqnarray}
\hspace*{-0.4cm}
B_{\mu\tau}
= - B_{\mu e} - B_{\mu\mu}
\simeq B_{e\mu}
\simeq 2{\rm Im}[S^{\ell *}_{\mu e} S^h_{\tau
e}]s_{23}c_{23}
\end{eqnarray}
from (\ref{relation-1}) and the unitarity relation.
We can derive the probability up to the second order
of two small parameters by using the unitarity
relation although we cannot directly calculate $S'_{\mu\tau}$
in the previous method.
In addition, the coefficients of $\sin 2\delta$ and
$\cos 2\delta$, $D$ and $E$, have an order of
\begin{eqnarray}
D = O(\alpha^2 s_{13}^2), \quad
E = O(\alpha^2 s_{13}^2)
\end{eqnarray}
in these three channels as derived from (\ref{D_mm}),
(\ref{E_mm}),
(\ref{D_tt}), (\ref{E_tt}), (\ref{D_mt}) and
(\ref{E_mt})
and are expected to be small.
Actually, these coefficients have the second order of
$S'_{\mu\tau}$, and the values are about $(0.02)^2
\simeq 0.0004$
from Figure 1 in the high energy region related with
long baseline experiments.
So we ignore them in the following section.

\subsection{Approximate Formulas in All Channels}

\hspace*{\parindent}
In this subsection, we present the approximate
formulas which are useful in arbitrary matter density profile.
Ignoring the higher order terms of $\alpha$ and
$s_{13}$ than the second order,
we can present the oscillation probabilities for all
channels with the amplitudes calculated in two generations.

At first, let us derive the approximate formulas for
$P(\nu_e \to \nu_\mu)$ and $P(\nu_e \to \nu_\tau)$.
The approximate formula for $P(\nu_e \to \nu_\mu)$
has already been derived in our previous paper
\cite{Takamura04}.
We only have to replace the amplitudes $S'_{\mu e}$
and $S'_{\tau e}$ in three generations into $S_{\mu
e}^\ell$ and $S_{\tau e}^h$ in two generations.
From (\ref{P_emu})-(\ref{C_etau}) and
(\ref{S_me})-(\ref{S_te}),
we obtain
\begin{eqnarray}
P(\nu_e \to \nu_\mu)
&=& A_{e\mu}\cos\delta + B_{e\mu}\sin\delta + C_{e\mu}
\label{P_em1} \\
A_{e\mu}
&\simeq& 2 {\rm Re}[{S}_{\mu e}^{\ell *} S^h_{\tau e}]
c_{23} s_{23},
\label{A_em1} \\
B_{e\mu}
&\simeq& - 2 {\rm Im}[{S}_{\mu e}^{\ell *} S^h_{\tau
e}] c_{23} s_{23},
\label{B_em1} \\
C_{e\mu} &\simeq& |S^\ell_{\mu e}|^2 c_{23}^2 +
|S^h_{\tau e}|^2 s_{23}^2,
\label{C_em1} \\
P(\nu_e \to \nu_\tau)
&=& A_{e\tau}\cos\delta + B_{e\tau}\sin\delta +
C_{e\tau}
\label{P_et1} \\
A_{e\tau}
&\simeq& - 2 {\rm Re}[{S}_{\mu e}^{\ell *} S^h_{\tau
e}] c_{23} s_{23},
\label{A_et1} \\
B_{e\tau}
&\simeq& 2 {\rm Im}[{S}_{\mu e}^{\ell *} S^h_{\tau e}]
c_{23} s_{23},
\label{B_et1} \\
C_{e\tau}
&\simeq& |S^\ell_{\mu e}|^2 s_{23}^2 + |S^h_{\tau
e}|^2 c_{23}^2,
\label{C_et1}.
\end{eqnarray}
Eqs. (\ref{P_em1})-(\ref{C_em1}) are the same as those
derived in
our previous paper \cite{Takamura04}.
Next, let us derive the approximate formulas for
$P(\nu_e \to \nu_e)$.
Using (\ref{S_ee}) directly, we obtain
\begin{eqnarray}
P(\nu_e \to \nu_e)
&=& C_{ee} = |S'_{ee}|^2 \\
&\simeq& |S_{ee}^\ell + S_{ee}^h-S^d_{ee}|^2
\label{C_ee1}.
\end{eqnarray}
On the other hand, we obtain
\begin{eqnarray}
P(\nu_e \to \nu_e)
&=& C_{ee} = 1- C_{e\mu} - C_{e\tau} \\
&\simeq& 1-|S^\ell_{\mu e}|^2 - |S^h_{\tau e}|^2
\label{C_ee2},
\end{eqnarray}
by using the unitarity relation.
This is a different approximate formula than
(\ref{C_ee1}).
Thus, there are two kinds of expressions (\ref{C_ee1})
and (\ref{C_ee2}) for $P(\nu_e \to \nu_e)$.
We checked numerically that the expression
(\ref{C_ee2}) has a better precision than the expression 
(\ref{C_ee1}).
Furthermore, the expression (\ref{C_ee2}) easily
reproduces the approximate formula derived with double 
expansion up to the second order of two small parameters 
in ref. \cite{Akhmedov04} (second order formula).
So we use the expression (\ref{C_ee2}) in the following.

Next, let us derive the approximate formula for
$P(\nu_\mu \to \nu_\tau)$.
At first we calculate the terms independent of the CP
phase in this calculation.
We can approximate
\begin{eqnarray}
C_{\mu\tau}
= |S'_{\mu \tau}|^2 s_{23}^4 + |S'_{\tau \mu}|^2
c_{23}^4
+ |S'_{\mu \mu}-S'_{\tau \tau}|^2c_{23}^2 s_{23}^2
\simeq |S^\ell_{\mu \mu} - S^h_{\tau \tau}|^2 c_{23}^2
s_{23}^2
\end{eqnarray}
from (\ref{C_mt}) and (\ref{S_mm})-(\ref{S_tt}),
where we ignore the terms proportional to
$|S'_{\mu\tau}|^2=O(\alpha^2 s_{13}^2)$.
This leads to the approximated probability as
\begin{eqnarray}
P(\nu_\mu \to \nu_\tau) &=& B_{\mu\tau}\sin\delta +
C_{\mu\tau}
\label{P_mt1} \\
B_{\mu\tau}
&\simeq& 2 {\rm Im}[{S}_{\mu e}^{\ell*} S^h_{\tau e}]
c_{23} s_{23},
\label{B_mt1} \\
C_{\mu\tau}
&\simeq& |S^\ell_{\mu \mu} - S^h_{\tau \tau}|^2
c_{23}^2 s_{23}^2
\label{C_mt1}.
\end{eqnarray}
Next, let us derive the approximate formulas for
$P(\nu_\mu \to \nu_\mu)$ and $P(\nu_\tau \to
\nu_\tau)$.
From (\ref{C_mm}) and (\ref{S_mm})-(\ref{S_tt}), we
obtain
\begin{eqnarray}
C_{\mu\mu}
&=& |S'_{\mu \mu} c_{23}^2+S'_{\tau \tau}s_{23}^2|^2
+ (|S'_{\mu \tau}|^2 + |S'_{\tau \mu}|^2) c_{23}^2
s_{23}^2 \\
&\simeq& |S_{\mu \mu}^\ell c_{23}^2+S_{\tau \tau}^h
s_{23}^2|^2
\label{C_mm2},
\end{eqnarray}
where we neglect the terms proportional to
$|S'_{\mu\tau}|^2=O(\alpha^2 s_{13}^2)$.
On the other hand, we obtain another expression by
using the
unitarity relation as
\begin{eqnarray}
C_{\mu\mu}
&=& 1- C_{\mu e} - C_{\mu\tau} \\
&\simeq& 1-|S^\ell_{\mu e}|^2 c_{23}^2 -|S^h_{\tau
e}|^2 s_{23}^2
- |S^\ell_{\mu \mu} - S^h_{\tau \tau}|^2 c_{23}^2
s_{23}^2
\label{C_mm3}
\end{eqnarray}
This seems to be different from (\ref{C_mm2}) at a
glance,
but we confirmed that (\ref{C_mm2}) and (\ref{C_mm3})
are the same expression by using the unitarity relation.
In the following, we use the expression (\ref{C_mm3})
for the reason that this easily reproduces the second
order formula and we can check the unitarity.
In the same way, $C_{\tau\tau}$ is given by
\begin{eqnarray}
C_{\tau\tau}
= 1- C_{e\tau} - C_{\mu\tau}
\simeq 1-|S^\ell_{\mu e}|^2 s_{23}^2 -|S^h_{\tau e}|^2
c_{23}^2
- |S^\ell_{\mu \mu} - S^h_{\tau \tau}|^2 c_{23}^2
s_{23}^2
\end{eqnarray}
from the unitarity relation.
From the result obtained in subsection 4.1,
the approximate formulas for
$P(\nu_\mu \to \nu_\mu)$ and $P(\nu_\tau \to\nu_\tau)$
are given by
\begin{eqnarray}
P(\nu_\mu \to \nu_\mu) &=& A_{\mu\mu}\cos\delta +
C_{\mu\mu}
\label{P_mm1} \\
A_{\mu\mu}
&\simeq& -2 {\rm Re}[{S}_{\mu e}^{\ell*} S^h_{\tau e}]
c_{23} s_{23},
\label{A_mm1} \\
C_{\mu\mu}
&\simeq& 1-|S^\ell_{\mu e}|^2 c_{23}^2 -|S^h_{\tau
e}|^2 s_{23}^2
- |S^\ell_{\mu \mu} - S^h_{\tau \tau}|^2 c_{23}^2
s_{23}^2
\label{C_mm1} \\
P(\nu_\tau \to \nu_\tau) &=& A_{\tau\tau}\cos\delta +
C_{\tau\tau}
\label{P_tt1} \\
A_{\tau\tau}
&\simeq& 2 {\rm Re}[{S}_{\mu e}^{\ell*} S^h_{\tau e}]
c_{23} s_{23},
\label{A_tt1} \\
C_{\tau\tau}
&\simeq& 1-|S^\ell_{\mu e}|^2 s_{23}^2 -|S^h_{\tau
e}|^2 c_{23}^2
- |S^\ell_{\mu \mu} - S^h_{\tau \tau}|^2 c_{23}^2
s_{23}^2
\label{C_tt1}.
\end{eqnarray}
These results are one of the main results of this
paper.
In all channels, we can present the probabilities
including the CP violation by using the amplitudes calculated in
two generations.
It is noted that the CP violating terms due to the
existence of three generations can be calculated from the two
generation amplitudes.

Next, let us compare the approximate formulas
(\ref{P_em1})-(\ref{C_et1}), (\ref{C_ee2}),
(\ref{P_mt1})-(\ref{C_mt1}) and
(\ref{P_mm1})-(\ref{C_tt1})
with the numerical calculations.
We take the PREM (Preliminary Reference Earth Model)
as the earth matter density profile and compare
the approximated values of all probabilities with
those calculated numerically.
We use the same parameters as those used in fig. 1
and $\sin 2 \theta_{23}=1$, $\delta=90^\circ$.
We set the baseline length, $L=6000~{\rm km}$ and
the energy region, $1~{\rm GeV} \leq E \leq 20~{\rm
GeV}$,
to include the high energy MSW resonance.

\begin{figure}[!htb]
\begin{tabular}{ccc}
$P(\nu_e \to \nu_\mu)$ & $P(\nu_e \to \nu_\tau)$ &
$P(\nu_\mu \to \nu_\tau)$ \\
\resizebox{50mm}{!}{\includegraphics{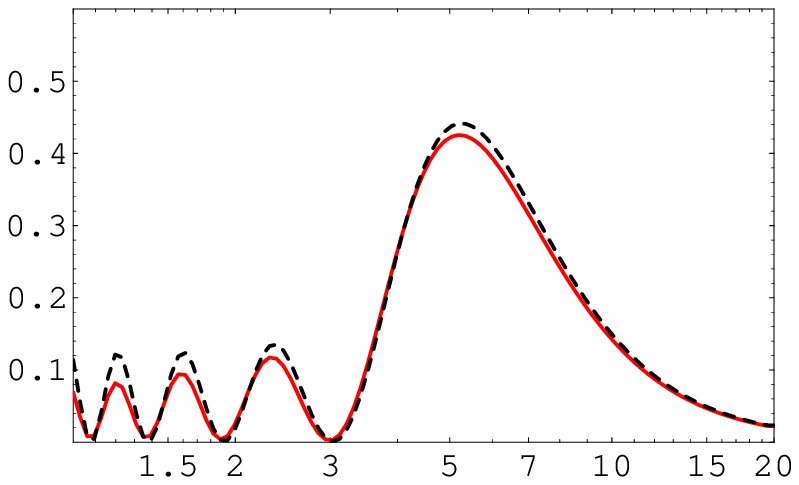}} &
\resizebox{50mm}{!}{\includegraphics{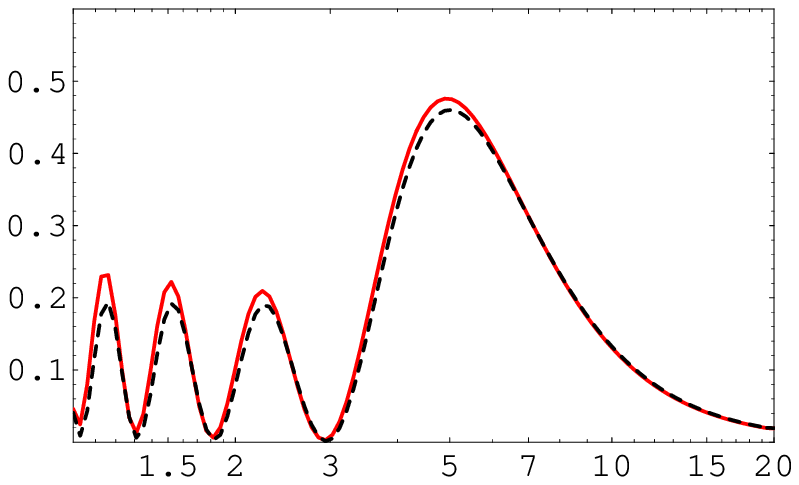}} &
\resizebox{50mm}{!}{\includegraphics{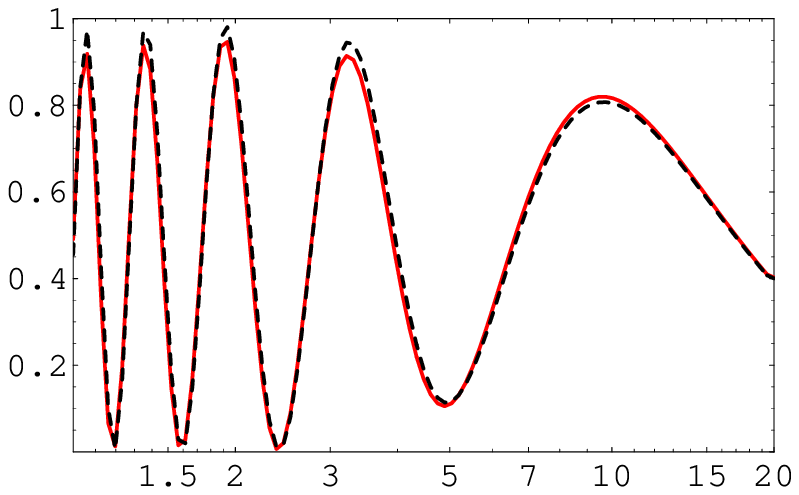}} \\
$P(\nu_e \to \nu_e)$ & $P(\nu_\mu \to \nu_\mu)$ &
$P(\nu_\tau \to \nu_\tau)$ \\
\resizebox{50mm}{!}{\includegraphics{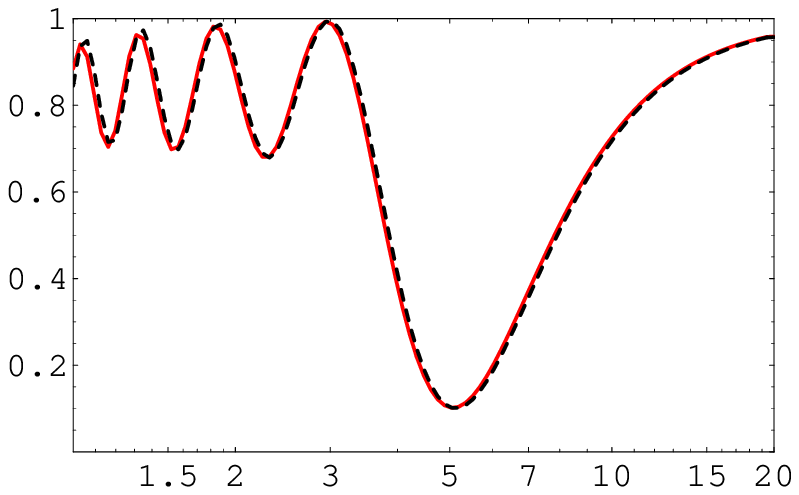}} &
\resizebox{50mm}{!}{\includegraphics{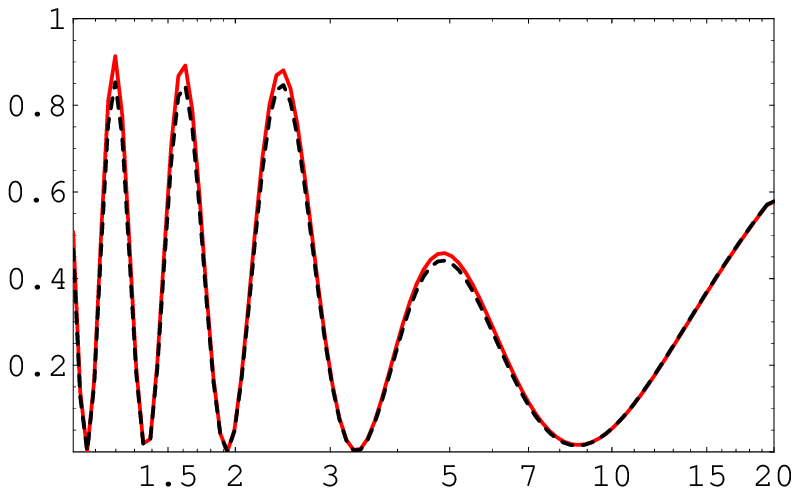}} &
\resizebox{50mm}{!}{\includegraphics{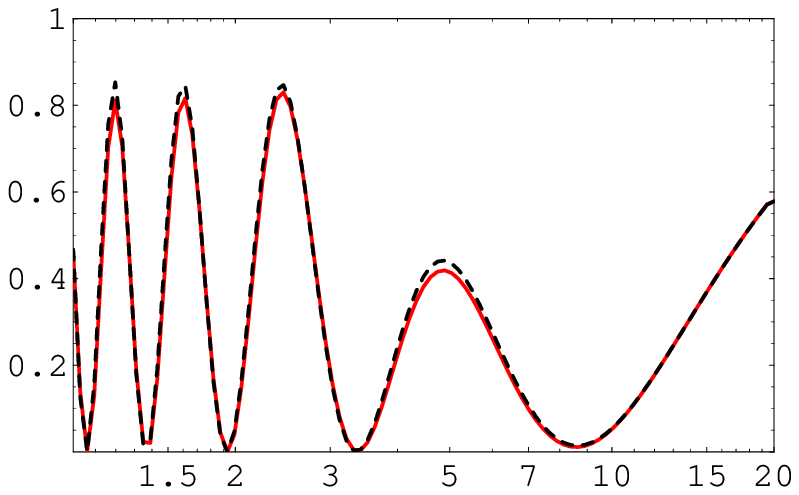}}
\end{tabular}
\caption{
Comparison of our approximate formulas with the
numerical calculation.
In these figures $P(\nu_\alpha \to \nu_\beta)$ is
plotted in order to compare our approximate formulas with the
numerical calculation.
The solid lines show the approximate probabilities and
the dashed lines show the numerical calculation of probabilities.
}
\end{figure}

We compare our approximate formulas with
the numerical calculation in figure 3.
One can see some features from this figure.
The approximated value of probabilities for $P(\nu_e
\to \nu_\mu),
P(\nu_e \to \nu_\tau)$ and $P(\nu_e \to \nu_e)$
coincide to the numerical values very well, 
on the other hand, the remaining three channels of
probabilities $P(\nu_\mu \to \nu_\tau), P(\nu_\mu \to \nu_\mu)$ 
and $P(\nu_\tau \to \nu_\tau)$ show a small difference
between the approximate and the numerical value.
However, the difference is not so large as in
figure 2 and as a first step the result is sufficiently
accurate.

\section{Comparison of Our Results with Second Order
Formulas}

\hspace*{\parindent}
In this section, we concretely calculate the
amplitudes by using the approximate formulas derived 
in the previous section
for the case of constant matter and show that simple
approximate formulas can be obtained.
Finally, we demonstrate that the approximate
formula derived with double expansion up to the second 
order of the two small parameters
(second order formulas) are largely different from the
exact values in the MSW resonance region under the condition
that the baseline length is longer than
the oscillation length.

\subsection{Approximate Formulas for Amplitudes}

\hspace*{\parindent}
In the previous section, we have given a method
for approximation of the probabilities in three generations by
amplitudes in two generations.
In this subsection, we use the constant matter density
profile in order to compare our method with other
method and
investigate the non-perturbative effects.
As seen from (\ref{P_em1})-(\ref{C_et1}),
(\ref{C_ee2}),
(\ref{P_mt1})-(\ref{C_mt1}) and
(\ref{P_mm1})-(\ref{C_tt1}),
we only have to calculate four kinds of amplitudes,
namely $S_{\mu e}^\ell, S_{\mu\mu}^\ell, S_{\tau e}^h$ and
$S_{\tau\tau}^h$.

The low-energy approximate formulas are obtained
by taking the
limit $s_{13} = 0$ and from
\begin{eqnarray}
H^\ell &=& O_{12}{\rm diag}(0,\Delta_{21},\Delta_{31})
O_{12}^T
+ {\rm diag}(a,0,0) \label{H-low} \\
&=& O^\ell_{12}{\rm
diag}(\lambda_1^\ell,\lambda_2^\ell,\Delta_{31})
(O_{12}^\ell)^T \label{H-low-diag}.
\end{eqnarray}
The effective masses $\lambda_i^\ell (i=1,2)$ and
the effective mixing angle $\theta_{12}^\ell$ are
determined
by the diagonalization of (\ref{H-low}) to
(\ref{H-low-diag}).
If we define the mass squared difference in matter as
$\Delta_{21}^\ell = \lambda_2^\ell-\lambda_1^\ell$,
we obtain the relation
\begin{eqnarray}
\frac{\Delta^\ell_{21}}{\Delta_{21}}
= \frac{\sin 2\theta_{12}}{\sin 2\theta^\ell_{12}}
= \sqrt{\left(
\cos 2\theta_{12}-\frac{a}{\Delta_{21}}
\right)^2 + \sin^2 2\theta_{12}}.
\end{eqnarray}
Therefore, the amplitude is calculated as
\begin{eqnarray}
S^\ell_{\mu e}
&=&
-i \sin 2\theta^\ell_{12} \sin
\frac{\Delta^\ell_{21}L}{2}
\exp\left(-i\frac{\Delta_{21} +a}{2}L\right)
\label{S_mel} \\
S^\ell_{\mu\mu}
&=&
\left(
\cos\frac{\Delta^\ell_{21}L}{2}
- i
\cos 2\theta^\ell_{12}\sin\frac{\Delta^\ell_{21}L}{2}
\right)\exp\left(-i\frac{\Delta_{21} +a}{2}L\right)
\label{S_mml}
\end{eqnarray}
by substituting (\ref{H-low-diag}) into (\ref{low-S}).
On the other hand, the approximate formulas in
high energy are obtained by taking the limit 
$\alpha = 0$ and we get
\begin{eqnarray}
H^h &=& O_{13}{\rm diag}(0,0,\Delta_{31})O_{13}^T
+ {\rm diag}(a,0,0) \label{H-high} \\
&=& O^h_{13}{\rm diag}(\lambda_1^h,0,\lambda_3^h)
(O^h_{13})^T \label{H-high-diag}.
\end{eqnarray}
The effective masses $\lambda_i^h (i=1,3)$ and the
effective mixing angle $\theta_{13}^h$ are determined by the
diagonalization
of (\ref{H-high}) to (\ref{H-high-diag}).
If we define the mass squared difference in matter as
$\Delta_{31}^h = \lambda_3^h-\lambda_1^h$,
we obtain the relation
\begin{eqnarray}
\frac{\Delta^h_{31}}{\Delta_{31}}
= \frac{\sin 2\theta_{13}}{\sin 2\theta^h_{13}}
= \sqrt{\left(
\cos 2\theta_{13}-\frac{a}{\Delta_{31}}
\right)^2 + \sin^2 2\theta_{13}}.
\end{eqnarray}
Accordingly, the amplitude can be calculated by
substituting (\ref{H-high-diag}) into (\ref{high-S})
as
\begin{eqnarray}
S^h_{\tau e}
&=&
-i \sin 2\theta^h_{13} \sin \frac{\Delta^h_{31}L}{2}
\exp\left(-i\frac{\Delta_{31} +a}{2}L \right)
\label{S_teh} \\
S^h_{\tau\tau}
&=&
\left(
\cos\frac{\Delta^h_{31}L}{2}
- i \cos 2\theta^h_{13}\sin\frac{\Delta^h_{31}L}{2}
\right)\exp\left(-i\frac{\Delta_{31} +a}{2}L
\label{S_tth}
\right).
\end{eqnarray}
As seen from (\ref{S_mel}) and (\ref{S_teh}),
$S_{\mu e}^\ell$ and $S_{\tau e}^h$ have simple forms,
but the expressions of $S_{\mu\mu}^\ell$ and
$S_{\tau\tau}^h$
are more complex than (\ref{S_mml}) and (\ref{S_tth}).

\subsection{Approximate Formulas for Probabilities}

\hspace*{\parindent}
In this subsection, we derive the approximate formulas
of
the oscillation probabilities in constant matter
by using the result of the previous section.

At first, let us consider the case of including
electron neutrino
in the initial or final state.
In this case, the probability for any channel can be
calculated
almost in the same way.
The probability $P(\nu_e \to \nu_e)$ is obtained by
substituting
(\ref{S_mel}) and (\ref{S_teh}) into
(\ref{C_ee2}) as
\begin{eqnarray}
P(\nu_e \to \nu_e)
= 1-\sin^2 2\theta^\ell_{12}
\sin^2 \frac{\Delta^\ell_{21}L}{2}
- \sin^2 2\theta^h_{13}
\sin^2 \frac{\Delta^h_{31}L}{2}.
\end{eqnarray}
The probability $P(\nu_e \to \nu_{\mu})$ is obtained
by substituting (\ref{S_mel}) and (\ref{S_teh})
into (\ref{A_em1})-(\ref{C_em1}) as
\begin{eqnarray}
P(\nu_e \to \nu_\mu)
&=& A_{e\mu}\cos\delta + B_{e\mu}\sin\delta + C_{e\mu}
\label{P_em2} \\
A_{e\mu}
&\simeq& \sin 2\theta^\ell_{12}\sin 2\theta_{23}\sin
2\theta^h_{13}
\sin \frac{\Delta^\ell_{21}L}{2}
\sin \frac{\Delta^h_{31}L}{2}
\cos \frac{\Delta_{32}L}{2} \label{A_em2} \\
B_{e\mu}
&\simeq& \sin 2\theta^\ell_{12}\sin 2\theta_{23}\sin
2\theta^h_{13}
\sin \frac{\Delta^\ell_{21}L}{2}
\sin \frac{\Delta^h_{31}L}{2}
\sin \frac{\Delta_{32}L}{2} \label{B_em2} \\
C_{e\mu} &\simeq& c_{23}^2\sin^2 2\theta^\ell_{12}
\sin^2 \frac{\Delta^\ell_{21}L}{2}
+ s_{23}^2\sin^2 2\theta^h_{13}
\sin^2 \frac{\Delta^h_{31}L}{2} \label{C_em2}.
\end{eqnarray}
The remaining probability $P(\nu_e \to \nu_\tau)$
can be calculated from the unitarity relation.
Next, let us calculate the probabilities for the
case, that not all electron neutrinos in the initial 
and final state are included.
Also in this case, the probability for any channel can
be calculated almost in the same way.
Accordingly, as an example, we calculate the
probability for muon neutrino to tau neutrino,
\begin{eqnarray}
P(\nu_\mu \to \nu_\tau)
&=& B_{\mu\tau}\sin\delta + C_{\mu\tau} \\
B_{\mu\tau}
&\simeq&
\sin 2\theta^\ell_{12}\sin 2\theta_{23}\sin
2\theta^h_{13}
\sin \frac{\Delta^\ell_{21}L}{2}
\sin \frac{\Delta^h_{31}L}{2}
\sin \frac{\Delta_{32}L}{2}
\label{B_mt-1} \\
C_{\mu\tau}
&\simeq&
|S_{\mu\mu}^\ell-S_{\tau\tau}^h|^2s_{23}^2c_{23}^2.
\end{eqnarray}
At first, we use the relations,
$\cos 2\theta_{12}^\ell = 2\cos^2 \theta_{12}^\ell-1$
and $\cos 2\theta_{13}^h = 1-2\sin^2 \theta_{13}^h$,
and we rewrite $S_{\mu\mu}^\ell$ and $S_{\tau\tau}^h$
as
\begin{eqnarray}
S_{\mu\mu}^\ell
&\simeq&
\left[
\exp
\left(i\frac{\Delta_{21}^\ell}{2}L\right)
- 2i
\cos^2\theta^\ell_{12}\sin\frac{\Delta^\ell_{21}L}{2}
\right]
\exp
\left(-i\frac{\Delta_{21} +a}{2}L
\right) \\
S_{\tau\tau}^h
&\simeq&
\left[
\exp
\left(
-i\frac{\Delta^h_{31}L}{2}
\right)
+2i \sin^2\theta^h_{13}\sin\frac{\Delta^h_{31}L}{2}
\right]
\exp
\left(-i\frac{\Delta_{31} +a}{2}L
\right).
\end{eqnarray}
Then, arranging $C_{\mu\tau}$ in the order of the
effective mixing angles $\cos \theta_{12}^\ell$ and
$\sin \theta_{13}^h$, we obtain
\begin{eqnarray}
C_{\mu\tau}^1
&=& \sin^2 2\theta_{23}
\sin^2
\frac{(\Delta_{21}^\ell+\Delta_{31}^h+\Delta_{32})
L}{4}
\label{C_mt-1} \\
C_{\mu\tau}^{2a}
&=& - 2\sin^2 2\theta_{23}\cos^2 \theta_{12}^\ell
\sin\frac{(\Delta_{21}^\ell+\Delta_{31}^h+\Delta_{32})L}{4}
\cos\frac{(\Delta_{21}^\ell-\Delta_{31}^h-\Delta_{32})L}{4}
\sin \frac{\Delta_{21}^\ell L}{2}
\label{C_mt-2a} \\
C_{\mu\tau}^{2b}
&=& - 2\sin^2 2\theta_{23}\sin^2 \theta_{13}^h
\sin\frac{(\Delta_{21}^\ell+\Delta_{31}^h+\Delta_{32})L}{4}
\cos\frac{(\Delta_{21}^\ell-\Delta_{31}^h+\Delta_{32})L}{4}
\sin \frac{\Delta_{31}^h L}{2}
\label{C_mt-2b} \\
C_{\mu\tau}^3
&=& \sin^2 2\theta_{23}
\cos^4\theta^\ell_{12}\sin^2\frac{\Delta^\ell_{21}L}{2}
+ \sin^2 2\theta_{23}\sin^4 \theta_{13}^h
\sin^2 \frac{\Delta_{31}^h L}{2} \nonumber \\
&+& 2\sin^2
2\theta_{23}\cos^2\theta^\ell_{12}\sin^2\theta^h_{13}
\sin\frac{\Delta^\ell_{21}L}{2}\sin\frac{\Delta^h_{31}L}{2}
\cos\frac{\Delta_{32}L}{2}
\label{C_mt-3}.
\end{eqnarray}
As we show in the following section, the reason of
arranging the terms like this is, because the second order
formulas can be easily derived.
In order to derive the second order formulas, it is
sufficient to use
$C_{\mu\tau}^1$, $C_{\mu\tau}^{2a}$ and $C_{\mu\tau}^{2b}$.
We can also calculate the other channels $P(\nu_\mu
\to \nu_\mu)$
and $P(\nu_\tau \to \nu_\tau)$ in the same way.
In a recent study, it was found that
the channels $P(\nu_\mu \to \nu_\mu)$ and
$P(\nu_\tau \to \nu_\tau)$ are largely affected by the
earth matter
in the long baseline
\cite{Choubey04, Gandhi04, Choudhury04}.

We can see from these expressions that the approximate
formulas are rather complex for the case not including
electron neutrino in the initial or final state.
We also understand from these formulas how matter
affects the probabilities.
Thus, the formulas are expected to be useful for
studying matter effects.

\subsection{Large Non-perturbative Effects of small
$\alpha$ and $s_{13}$}

\hspace*{\parindent}
In this subsection, we compare the approximate
formulas obtained in the previous subsection with the second
order formulas numerically and it is shown that
the latter have a large difference from the numerical
value in the MSW resonance region.

The second order formulas are approximated by the main
terms of the expansion and are widely used by many authors
for their simplicity.
In refs. \cite{Cervera00,Freund01},
the formula for $P(\nu_e \to \nu_\mu)$ has been
derived
and later on all probabilities were presented in
ref. \cite{Akhmedov04}.
As examples, the probabilities, $P(\nu_e \to \nu_\mu)$
and $P(\nu_\mu \to \nu_\tau)$, are taken.
For the other channels of probabilities, similar
expressions have been obtained.
In all channels similar results were obtained
from comparison with numerical calculations.
The second order formula for $P(\nu_e \to \nu_\mu)$ is
given by
\begin{eqnarray}
P(\nu_e \to \nu_\mu)
&=& A_{e\mu} \cos\delta
+ B_{e\mu} \sin\delta
+ C_{e\mu}, \label{double-P_em} \\
A_{e\mu}
&\simeq&
\alpha s_{13}\sin 2\theta_{12}\sin 2\theta_{23}
\frac{2\Delta_{31}^2}{a(\Delta_{31}-a)}
\sin \frac{aL}{2}\sin \frac{(\Delta_{31}-a)L}{2}
\cos \frac{\Delta_{32}L}{2}
\label{double-A_em} \\
B_{e\mu}
&\simeq&
\alpha s_{13}\sin 2\theta_{12}\sin 2\theta_{23}
\frac{2\Delta_{31}^2}{a(\Delta_{31}-a)}
\sin \frac{aL}{2}\sin \frac{(\Delta_{31}-a)L}{2}
\sin \frac{\Delta_{32}L}{2}
\label{double-B_em} \\
C_{e\mu}
&\simeq&
\alpha^2c_{23}^2\sin^2 2\theta_{12}
\frac{\Delta_{31}^2}{a^2}
\sin^2 \frac{aL}{2}
+ s_{13}^2s_{23}^2
\frac{4\Delta_{31}^2}{(\Delta_{31}-a)^2}
\sin^2 \frac{(\Delta_{31}-a)L}{2}
\label{double-C_em}.
\end{eqnarray}
Comparing our approximate formulas
(\ref{A_em2})-(\ref{C_em2}) with
the second order formulas
(\ref{double-A_em})-(\ref{double-C_em}),
each term corresponds one by one.
Actually, the second order formulas
(\ref{double-A_em})-(\ref{double-C_em}) are derived by
expanding our approximate formulas (\ref{A_em2})-(\ref{C_em2})
up to the second order in $\alpha$ and $s_{13}$
\cite{Takamura04}.
Next, the second order formula for $P(\nu_\mu \to \nu_\tau)$ 
which has been already derived in ref. \cite{Akhmedov04}
is
\begin{eqnarray}
P(\nu_\mu \to \nu_\tau)
&=& A_{\mu\tau}\cos\delta + B_{\mu\tau}\sin\delta +
C_{\mu\tau} \\
A_{\mu\tau}
&\simeq&
\alpha s_{13}\sin^2 2\theta_{23}\sin 2\theta_{12}
\cos 2\theta_{23}\frac{2\Delta_{31}}{\Delta_{31}-a}
\sin\frac{\Delta_{31}L}{2}
\label{double-A_mt} \nonumber \\
&\times&
\left[
\frac{a}{\Delta_{31}}
\sin\frac{\Delta_{31}L}{2}
-\frac{\Delta_{31}}{a}
\sin\frac{aL}{2}
\cos\frac{(\Delta_{31}-a)L}{2}
\right] \\
B_{\mu\tau}
&\simeq&
\alpha s_{13}\sin 2\theta_{12}\sin 2\theta_{23}
\frac{2\Delta_{31}^2}{a(\Delta_{31}-a)}
\sin \frac{aL}{2}\sin \frac{(\Delta_{31}-a)L}{2}
\sin \frac{\Delta_{32}L}{2},
\end{eqnarray}
and $C_{\mu\tau}$ is given by
\begin{eqnarray}
C_{\mu\tau}
&\simeq&
\sin^2 2\theta_{23}\sin^2 \frac{\Delta_{31}L}{2}
\nonumber \\
&-& \alpha \sin^2 2\theta_{23}\cos^2 \theta_{12}
\left(\frac{\Delta_{31}L}{2}\right)\sin\Delta_{31}L
+ \alpha^2 \sin^2 2\theta_{23}\cos^4 \theta_{12}
\left(\frac{\Delta_{31}L}{2}\right)^2
\cos\Delta_{31}L
\nonumber \\
&-& \alpha^2 \sin^2 2\theta_{23}\sin^2 2\theta_{12}
\left(\frac{\Delta_{31}}{2a}\right)
\left[
\sin
\frac{\Delta_{31}L}{2}
\cos\frac{(\Delta_{31}-a)L}{2}
\sin\frac{aL}{2}
\left(
\frac{\Delta_{31}}{a}
\right)
-\frac{\Delta_{31}L}{4}\sin(\Delta_{31}L)
\right]
\nonumber \\
&-& s_{13}^2\sin^2
2\theta_{23}\frac{2\Delta_{31}}{\Delta_{31}-a}
\left[
\sin\frac{\Delta_{31}L}{2}
\cos\frac{aL}{2}
\sin \frac{(\Delta_{31}-a)L}{2}
\left(
\frac{\Delta_{31}}{\Delta_{31}-a}
\right)
-\frac{aL}{4}\sin(\Delta_{31}L)
\right]
\label{double-C_mt}.
\end{eqnarray}
In the next section we show that this formula
(\ref{double-C_mt}) can be also derived from our formulas 
(\ref{C_mt-1})-(\ref{C_mt-3}).
It is noted that the second order formula (\ref{double-C_mt})
for $C_{\mu\tau}$ is rather complex.
Furthermore comparing our approximate formula 
(\ref{B_mt-1})-(\ref{C_mt-3}) with the second order formula
(\ref{double-A_mt})-(\ref{double-C_mt}), we see that
$A_{\mu\tau}$ is not included in our formula.
The reason is, that $A_{\mu\tau}=0$ in the case of
maximal mixing angle $\sin 2\theta_{23} = 1$ and there 
is no way of calculating this by the
method described in this paper.
If we consider $\cos2\theta_{23}$ as a small parameter
like $\alpha$ and $s_{13}$, this $A_{\mu\tau}$ has the
magnitude of $O(\alpha s_{13} \cos 2\theta_{23})$.
Therefore, $A_{\mu\tau}$ is proportional to the
third order of small parameters and is expected to 
be neglectable.
This means that our formula is not largely affected by
the error due to this term, which cannot be derived 
from our method.
However, as this error affects the precision measurement
of $\sin \theta_{23}$ by the atmospheric neutrino
experiments in future,
the improvement of this point is important future work.
The formulas for the other channels are given in ref.
\cite{Akhmedov04}.
The second order formulas are effective under the
following two conditions.

The first one is for the neutrino energy and is given
by
\begin{eqnarray}
E \gg 0.45 \: {\rm GeV}
\left(
\frac{\Delta m_{21}^2}{10^{-4}\:{\rm eV^2}}
\right)
\left(
\frac{3\:{\rm g/cm^3}}{\rho}
\right) \label{energy-condition}.
\end{eqnarray}
The second one is for the baseline length and is given
by
\begin{eqnarray}
L \ll 8000 \: {\rm km}
\left(
\frac{E}{{\rm GeV}}
\right)
\left(
\frac{10^{-4}\:{\rm eV^2}}{\Delta m_{21}^2}
\right) \label{baseline-condition}.
\end{eqnarray}
These conditions come from the utilization of
perturbative expansion
on the two small parameters.
The detailed discussion are given in \cite{Freund01}.
These approximate formulas are used for the purpose 
of understanding of the results obtained by numerical 
calculations \cite{Huber02,Huber04}.
However, as shown in the next figure, these formulas
have large difference from the true value in the MSW 
resonance region, which is considered to be the most 
important region.

Next, let us compare our formulas
(\ref{P_em2})-(\ref{C_mt-2b})
with the second order formulas
(\ref{double-P_em})-(\ref{double-C_mt})
in all channels by numerical calculation.
In order to see the magnitude of the error, we also
compare two kinds of approximate formulas with the exact
values.
We set the baseline length as $L=6000~{\rm km}$,
where the MSW effect becomes large, and the energy
region as $1~{\rm GeV} \leq E \leq 20~{\rm GeV}$, where the MSW
resonance energy is included.
Furthermore, the second order formulas are derived
only in the case of constant matter, so we choose the
average density
$\rho= 3.91~{\rm g/cm^3}$ of the earth calculated by
the PREM.
Note that two conditions
(\ref{energy-condition}) and
(\ref{baseline-condition})
are satisfied in these region.

\begin{figure}[!htb]
\begin{tabular}{ccc}
$P(\nu_e \to \nu_\mu)$ & $P(\nu_e \to \nu_\tau)$ &
$P(\nu_\mu \to \nu_\tau)$ \\
\resizebox{50mm}{!}{\includegraphics{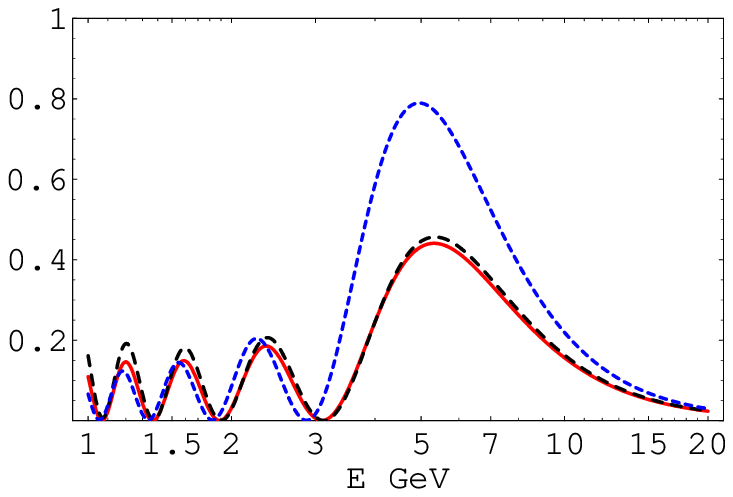}}
&
\resizebox{50mm}{!}{\includegraphics{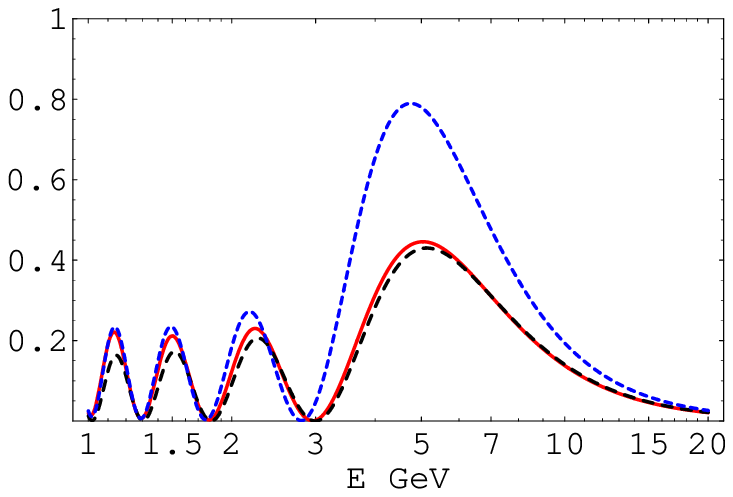}}
&
\resizebox{50mm}{!}{\includegraphics{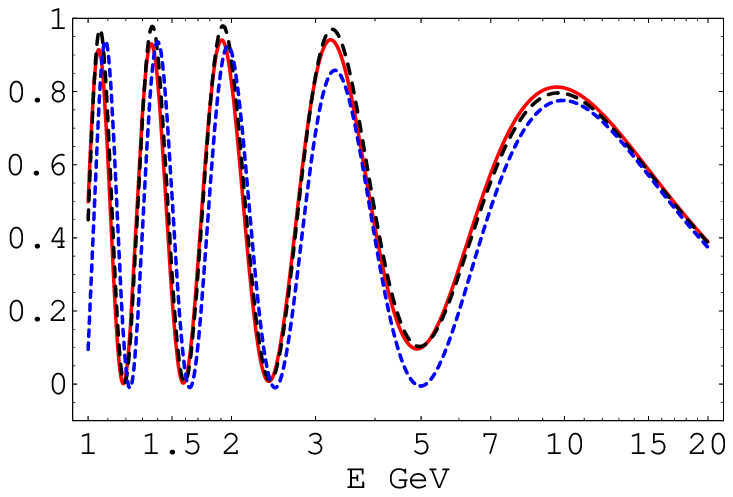}}
\\
$P(\nu_e \to \nu_e)$ & $P(\nu_\mu \to \nu_\mu)$ &
$P(\nu_\tau \to \nu_\tau)$ \\
\resizebox{50mm}{!}{\includegraphics{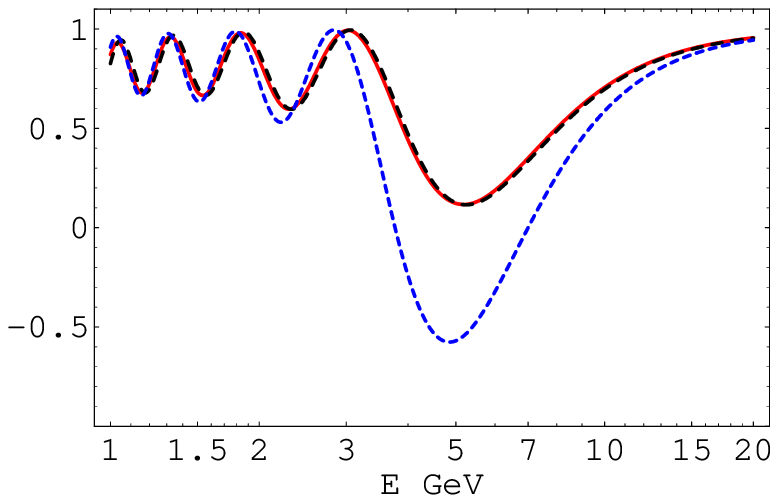}}
&
\resizebox{50mm}{!}{\includegraphics{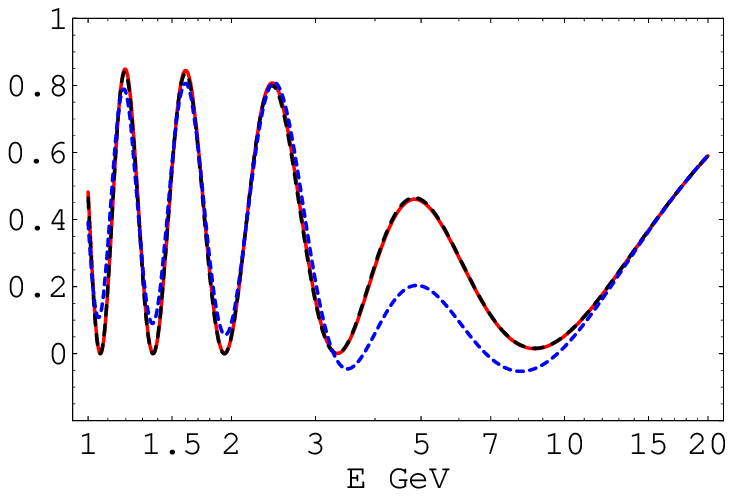}}
&
\resizebox{50mm}{!}{\includegraphics{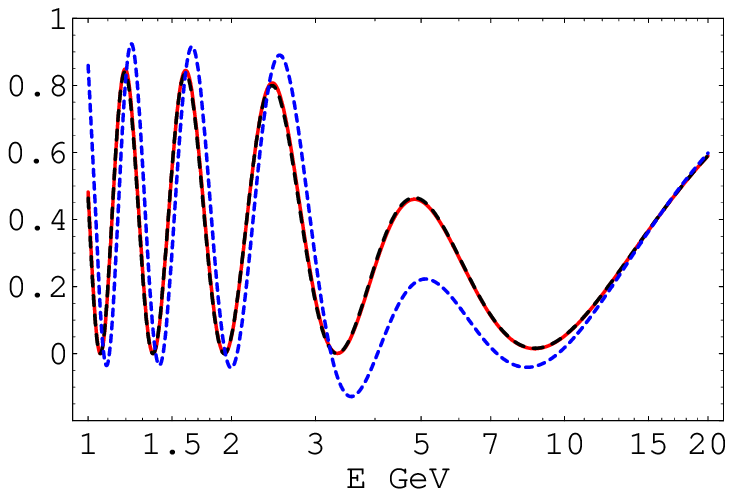}}
\end{tabular}
\caption{
Comparison of our approximate formulas with the
second order formulas
and in addition with the numerical calculations.
The solid, dashed and dotted lines show the
probabilities
in our approximate formulas, those in the numerical
calculation and
those in the second order formulas, respectively.
}
\end{figure}
We compare the probabilities calculated by our
approximate formulas
in all channels with those by the second order
formulas in addition to numerical calculation in figure 4.
One can see the following points from this figure.
The second order formulas show large differences
from the numerical values around 5 GeV, where
the high energy MSW resonance occurs.
In other energy regions they are in good coincidence.
The value of $P(\nu_e \to \nu_e)$ has the largest
difference,
the probabilities $P(\nu_e \to \nu_\mu)$ and $P(\nu_e
\to \nu_\tau)$
have the next largest difference,
the values of $P(\nu_\mu \to \nu_\mu)$ and $P(\nu_\tau
\to \nu_\tau)$
have also significantly large difference, but only the
probability
$P(\nu_\mu \to \nu_\tau)$ has a small difference.
In addition, these figures show that
the difference between the second order formulas
and the numerical calculation exists even in the two
applicable regions (\ref{energy-condition}) and
(\ref{baseline-condition}).
Although we do not show a figure, the difference
between our approximate formulas (\ref{P_em2})-(\ref{C_mt-2b})
and the second order formulas
(\ref{double-P_em})-(\ref{double-C_mt})
become more clear out of the two applicable regions
(\ref{energy-condition}) and
(\ref{baseline-condition}).
The reason is that our approximate formulas
(\ref{P_em2})-(\ref{C_mt-2b}) are applicable even for
the case that the condition (\ref{energy-condition})
or (\ref{baseline-condition}) for energy and baseline
length does not hold,
as confirmed from the comparison with the exact
numerical calculation.
However, the second order formulas are good
approximations, when
the neutrino energy is not near the resonance energy,
even if the baseline length is long.

\section{Non-perturbative Effects of Small Parameters
$\Delta m_{21}^2/\Delta m_{31}^2$ and
$\sin\theta_{13}$}

\hspace*{\parindent}
In this section, we investigate the reason for
the difference
between the second order formula,which contains
the approximation with double expansion up to the 
second order of two small parameters,
and the numerical calculation around the MSW resonance
region as explained in the previous subsection.

We discuss the non-perturbative effects of small
mixing angle more detailed than in section 2.

\subsection{Derivation of the Second Order Formulas}

\hspace*{\parindent}
In this subsection, we investigate how the second
order formulas
are approximated expanding on $\alpha=\Delta
m_{21}^2/\Delta m_{31}^2$
In the previous paper, we have discussed the
probability
$P(\nu_e \to \nu_\mu)$, so we calculate the second
order formula for $P(\nu_\mu \to \nu_\tau)$ here.
The method of calculation is basically the same
but the calculation itself becomes slightly complex,
because we need to calculate the effective mass and the
effective mixing angle up to the second order of $\alpha$ 
and $s_{13}$ in the case of $P(\nu_\mu \to \nu_\tau)$.
In this point, the calculation is not straightforward
compared with that of $P(\nu_e \to \nu_\mu)$ but the method 
of approximation is the same.
Note that $C_{\mu\tau}^1$ in (\ref{C_mt-1}) 
does not include the
effective mixing angle.
For this reason, it is sufficient to expand the
effective mixing angle up to the
zeroth order, but we need to expand the effective mass
up to the second order to calculate the probability 
up to the second order in $\alpha$ and $s_{13}$,
The effective mixing angle is expanded up to the
zeroth order as
\begin{eqnarray}
\cos \theta_{12}^\ell
&\simeq& \frac{1}{2}\sin 2\theta_{12}^\ell
\simeq \alpha\sin 2\theta_{12}\frac{\Delta_{31}}{2a}
\label{mixing-12} \\
\sin \theta_{13}^h
&\simeq& \frac{1}{2}\sin 2\theta_{13}^h
\simeq s_{13}\frac{\Delta_{31}}{\Delta_{31}-a}
\label{mixing-13}
\end{eqnarray}
and the effective mass is expanded up to the second
order as
\begin{eqnarray}
\Delta^\ell_{21} &\simeq& a- \alpha\cos
2\theta_{12}\Delta_{31}
+ \alpha^2\sin^2 2\theta_{12}\frac{\Delta_{31}^2}{2a}
\label{mass-12} \\
\Delta^h_{31} &\simeq& \Delta_{31}- a
+ s_{13}^2\frac{2\Delta_{31}a}{\Delta_{31}-a}.
\label{mass-13}
\end{eqnarray}
Here, we should emphasize the following points.
Eqs. (\ref{mixing-12}) and (\ref{mass-12}) obtained by
the expansion in $\alpha$ diverge in the vacuum limit
$a \to 0$ and eqs. (\ref{mixing-13}) and (\ref{mass-13})
obtained by the expansion in $s_{13}$ diverge in the 
high energy MSW resonance limit $a \to \Delta_{31}$.
As shown in the following, these divergences cancel
and the probability has a finite value.
Expanding $C_{\mu\tau}^1$ in (\ref{C_mt-1}) up to 
the second order, we obtain
\begin{eqnarray}
C_{\mu\tau}^1
&\simeq& C_{\mu\tau}^{1a} + C_{\mu\tau}^{1b}
+ C_{\mu\tau}^{1c}+ C_{\mu\tau}^{1d}+ C_{\mu\tau}^{1e}
\\
C_{\mu\tau}^{1a}
&=&
\sin^2 2\theta_{23}\sin^2 \frac{\Delta_{31}L}{2} \\
C_{\mu\tau}^{1b}
&=&
- \alpha \sin^2 2\theta_{23}\cos^2 \theta_{12}
\left(\frac{\Delta_{31}L}{2}\right)\sin\Delta_{31}L \\
C_{\mu\tau}^{1c}
&=&
\alpha^2 \sin^2 2\theta_{23}\cos^4 \theta_{12}
\left(\frac{\Delta_{31}L}{2}\right)^2
\cos\Delta_{31}L \\
C_{\mu\tau}^{1d}
&=&
\alpha^2 \sin^2 2\theta_{23}
\sin^2 2\theta_{12}
\left(\frac{\Delta_{31}^2L}{8a}\right)\sin\Delta_{31}L
\\
C_{\mu\tau}^{1e}
&=&
s_{13}^2 \sin^2 2\theta_{23}
\left(\frac{a\Delta_{31}}{\Delta_{31}-a}\frac{L}{2}
\right)\sin\Delta_{31}L.
\end{eqnarray}
We also expand $C_{\mu\tau}^{2a}$ in (\ref{C_mt-2a}) and
$C_{\mu\tau}^{2b}$ in (\ref{C_mt-2b}) as
\begin{eqnarray}
C_{\mu\tau}^{2a}
&\simeq& - 2\alpha^2 \sin^2 2\theta_{23}\sin^2
2\theta_{12}
\left(\frac{\Delta_{31}}{2a} \right)^2
\sin\frac{\Delta_{31}L}{2}
\cos\frac{(\Delta_{31}-a)L}{2}
\sin \frac{aL}{2} \\
C_{\mu\tau}^{2b}
&\simeq& - 2s_{13}^2\sin^2 2\theta_{23}
\left(\frac{\Delta_{31}}{\Delta_{31}-a}\right)^2
\sin\frac{\Delta_{31}L}{2}
\cos\frac{aL}{2}\sin \frac{(\Delta_{31}-a) L}{2}.
\end{eqnarray}
Finally, we obtain (\ref{double-C_mt}) arranging these
result order by order.

Here, let us consider the applicable region of the
second order formulas.
$C_{\mu\tau}^{1d}$ diverges in the limit $a \to 0$ and
$C_{\mu\tau}^{1e}$ diverges in the limit $a \to
\Delta_{31}$.
$C_{\mu\tau}^{2a}$ also diverges in the limit $a \to
0$ and
$C_{\mu\tau}^{2b}$ diverges in the limit $a \to
\Delta_{31}$.
The divergences in $a \to 0$ and in $a \to
\Delta_{31}$
come from the expansion of the effective masses
(\ref{mass-12}) and (\ref{mass-13}) respectively.
It seems that the second order formulas do not reduce
to those in vacuum due to the divergence
in $a \to 0$ and furthermore do not reduce to those in
the high energy MSW resonance point due to the divergence 
in $a \to \Delta_{31}$.
However, when we consider the pair
\begin{eqnarray}
C_{\mu\tau}^{1d} + C_{\mu\tau}^{2a}
= - \alpha^2 \sin^2 2\theta_{23}\sin^2 2\theta_{12}
\frac{\Delta_{31}}{2a}
\left[
\sin
\frac{\Delta_{31}L}{2}
\cos\frac{(\Delta_{31}-a)L}{2}
\sin\frac{aL}{2}
\left(
\frac{\Delta_{31}}{a}
\right)
-\frac{\Delta_{31}L}{4}\sin(\Delta_{31}L)
\right],
\end{eqnarray}
the divergence in $a \to 0$ cancel and the value
converges.
The obtained finite value is given by
\begin{eqnarray}
\lim_{a \to 0}(C_{\mu\tau}^{1d} + C_{\mu\tau}^{2a})
= - \alpha^2 \sin^2 2\theta_{23}\sin^2 2\theta_{12}
\frac{1}{2}\left(\frac{\Delta_{31}L}{2}\right)^2
\sin^2 \frac{\Delta_{31}L}{2}.
\end{eqnarray}
Before we expand, $C_{\mu\tau}^1$ and $C_{\mu\tau}^2$
have finite values in the limit $a \to 0$ and $a \to
\Delta_{31}$.
However, the divergence appears in expansion of
$\alpha$ and $s_{13}$.
The cancellation of these divergences occurs between
$C_{\mu\tau}^{1d}$ and $C_{\mu\tau}^{2a}$.
This means that the cancellation occurs between the
different terms
and result in finite values, respectively, at first,
which is an interesting result.
Considering the pair as
\begin{eqnarray}
C_{\mu\tau}^{1e} + C_{\mu\tau}^{2b}
= - s_{13}^2\sin^2
2\theta_{23}\frac{2\Delta_{31}}{\Delta_{31}-a}
\left[
\sin\frac{\Delta_{31}L}{2}
\cos\frac{aL}{2}
\sin \frac{(\Delta_{31}-a)L}{2}
\left(
\frac{\Delta_{31}}{\Delta_{31}-a}
\right)
-\frac{aL}{4}\sin(\Delta_{31}L)
\right],
\end{eqnarray}
the divergence in the limit $a \to \Delta_{31}$
cancels and
the value converges.
The finite value is given by
\begin{eqnarray}
\lim_{a \to \Delta_{31}}(C_{\mu\tau}^{1e} +
C_{\mu\tau}^{2b})
= s_{13}^2\sin^2 2\theta_{23}
\left(
\frac{\Delta_{31}L}{2}
\right)
\left[
(\Delta_{31}L)\sin^2\frac{\Delta_{31}L}{2}
- \sin (\Delta_{31}L)
\right].
\end{eqnarray}
The cancellation of these divergences occurs between
the different terms $C_{\mu\tau}^{1e}$ and
$C_{\mu\tau}^{2b}$,
which is also a remarkable result.

We have shown that the second order formulas
have finite values in the limit $a \to 0$ and $a \to
\Delta_{31}$,
but it is not always the same as that in the numerical
calculation.
Actually, the difference in fig. 4 in the limit $a \to
\Delta_{31}$,
shows that the second order formulas have finite
values but they are not in accordance with those
in the numerical calculation.
In order to study this, we compare the three
quantities,
the numerical calculation, our formulas and the second
order formulas.
We can learn the differences mainly
in the vacuum limit $a \to 0$ and the high energy MSW
resonance limit $a \to \Delta_{31}$ from the comparison.
\par
At first, let us consider the vacuum limit $a \to 0$.
Furthermore, to simplify the discussion, we consider
the case of $s_{13} \to 0$.
The second order formulas in the limits $a \to 0$ and
$s_{13} \to 0$
are given by
\begin{eqnarray}
\lim_{a, s_{13} \to 0}C_{\mu\tau}^{\rm (double)}
&=& \sin^2 2\theta_{23}
\sin^2 \frac{\Delta_{31} L}{2}
- \alpha \sin^2 2\theta_{23}\cos^2 \theta_{12}
\left(\frac{\Delta_{31}L}{2}\right)\sin\Delta_{31}L
\nonumber \\
&+& \alpha^2 \sin^2 2\theta_{23}\cos^4 \theta_{12}
\left(\frac{\Delta_{31}L}{2}\right)^2
\cos\Delta_{31}L \nonumber \\
&-& \alpha^2 \sin^2 2\theta_{23}\sin^2 2\theta_{12}
\frac{1}{2}\left(\frac{\Delta_{31}L}{2}\right)^2
\sin^2 \frac{\Delta_{31}L}{2} \label{double-C_mt-low}.
\end{eqnarray}
Next, taking the limit $a \to 0$ and $s_{13} \to 0$
in our formulas, we obtain
\begin{eqnarray}
\lim_{ a, s_{13} \to 0}C_{\mu\tau}^{\rm (exact)}
&=& \sin^2 2\theta_{23}
\sin^2 \frac{\Delta_{31} L}{2} \nonumber \\
&-& 2\sin^2 2\theta_{23}\cos^2 \theta_{12}
\sin\frac{\Delta_{31}L}{2}
\cos\frac{(\Delta_{21}-\Delta_{31})L}{2}
\sin \frac{\Delta_{21} L}{2} \label{our-C_mt-low}.
\end{eqnarray}
Expanding the oscillating part of (\ref{our-C_mt-low})
in our formula, it leads to (\ref{double-C_mt-low})
obtained from the second order formula.
The condition for the expansion on the oscillating
part for sufficiently good approximation is
\begin{eqnarray}
L < \frac{2}{\Delta_{21}}.
\end{eqnarray}

Next, let us consider the high energy MSW resonance
limit $a \to \Delta_{31}$.
In order to simplify the discussion, we take the
high energy
MSW resonance limit $a \to \Delta_{31}$ under the
condition $\alpha \to 0$.
In the high energy MSW resonance limit of the second
order formula, we obtain
\begin{eqnarray}
\lim_{a \to \Delta_{31}, \alpha \to 0}C_{\mu\tau}^{\rm
(double)}
&=& \sin^2 2\theta_{23}
\sin^2 \frac{\Delta_{31} L}{2} \nonumber \\
&+& s_{13}^2\sin^2 2\theta_{23}
\left(
\frac{\Delta_{31}L}{2}
\right)
\left[
(\Delta_{31}L)\sin^2\frac{\Delta_{31}L}{2}
- \sin (\Delta_{31}L)
\right] \label{double-C_mt-high}.
\end{eqnarray}
Next, taking the limit $a \to \Delta_{31}$ and $\alpha
\to 0$ in our formulas, we obtain
\begin{eqnarray}
\lim_{a \to \Delta_{31}, \alpha \to 0}C_{\mu\tau}^{\rm
(exact)}
&=& \sin^2 2\theta_{23}
\sin^2 \frac{(1+s_{13})\Delta_{31}L}{2}
\nonumber \\
&-& \sin^2 2\theta_{23}(1-s_{13}^2)
\sin\frac{(1+s_{13})\Delta_{31}L}{2}
\cos\frac{(1-s_{13})\Delta_{31}L}{4}
\sin (s_{13}\Delta_{31}L). \label{our-C_mt-high}
\end{eqnarray}
By expanding the oscillating part obtained from
our formula (\ref{our-C_mt-high}), it is shown that 
this coincides with that from the second order formula
(\ref{double-C_mt-high}).
The condition for the expansion of the oscillating
part for a sufficient
approximation is given by
\begin{eqnarray}
L < \frac{2}{s_{13}\Delta_{31}} = \frac{2}{s_{13}a}.
\end{eqnarray}
If the baseline length is shorter than that obtained
from
above condition, the second order formula becomes a
good approximation.
We obtain the following results about the perturbative
expansion
on the small parameters $\alpha$ and $s_{13}$.
\begin{enumerate}
\item
The perturbative expansion in $\alpha$ actually
corresponds to
the expansion in $\Delta_{21}/a$.
This constrains the applicable energy for the
approximate formulas.
If we expand in the parameter $\Delta_{21}/a$,
the effective mass $\Delta^\ell_{21}$ and the
effective mixing angle
$\sin 2\theta^\ell_{12}$ diverge in the vacuum limit
$a \to 0$.
However, these divergences cancel out each other
in the calculation of the
oscillation probability.
Thus, the probability has a finite value, but the
value largely
differs from the numerical calculation in low-energy.
The magnitude of this difference becomes large and
serious in the case of
small mixing angles and in low-energy long baseline
experiments.
\item
If we expand in the small mixing angle $s_{13}$,
the effective mass $\Delta^h_{31}$ and the effective
mixing angle
$\sin 2\theta^h_{13}$ diverge in the MSW resonance
energy limit $a \to \Delta_{31}$.
However, these divergences also cancel each other
out in the calculation of the
oscillation probability.
Thus, the probability has a finite value, but the
value largely differs from the numerical calculation in the
high-energy MSW resonance region.
This means that the second order formulas cannot be
used in the high energy MSW resonance region.
\end{enumerate}
In two generations, we can calculate the oscillation
probabilities
exactly by solving the second order equation.
So, we do not need the perturbative expansion.
On the other hand, the construction of the approximate
formulas applicable to arbitrary matter density profile 
is very difficult in three generations.
Therefore, we need to expand on the small parameters
$\alpha$ and $s_{13}$.

\subsection{Discussion}

\hspace*{\parindent}
We have shown that the double expansion formulas
up to the second order in the two small parameters
$\alpha$ and $s_{13}$ does not give a good
approximation
in the MSW resonance region.
This is because the coefficients of the small
parameters have
large values in the MSW resonance region.
In this subsection, let us discuss some methods
proposed up to present to solve this problem.
The Hamiltonian $H'$ is written by four parameters.
The two parameters $(\Delta m^2_{21},\theta_{12})$
control the physics mainly in the low-energy region 
and the other two parameters
$(\Delta m^2_{31},\theta_{13})$ control the physics
mainly in the high-energy region.
In other words, the magnitude of $\alpha$ determines
low-energy phenomena and the magnitude of $s_{13}$ determines
high-energy phenomena.
Both of these parameters are very small but the energy
region, where the expansion converges, is different.
This means that we need to treat the applicable energy
region
carefully when we expand on these two parameters.
There are several methods in order to take into
account the higher order terms of $\alpha$ and $s_{13}$ 
for example

\begin{enumerate}
\item exact formulas in constant matter density
profile
\item reduction formulas taking into account the two
generation
part exactly
\end{enumerate}

In the first method, there does not exist any
error generated from the perturbative expansion, 
because of the exact treatment
of both $\alpha$ and $s_{13}$ \cite{Kimura02}.
Furthermore, non-perturbative effects can be easily
investigated by using these exact formulas.
The second method was introduced in our previous
paper \cite{Takamura04}.
In this method, we try to include the higher order
terms of $\alpha$ and $s_{13}$ partially, except for 
the terms including the product of two small parameters.
This method includes the higher order terms of
$\alpha$ and $s_{13}$ and is simply and applicable 
even in the case of arbitrary matter density
profile \cite{Takamura04,Kimura04}.
Although this method uses only the second order
approximation of the amplitude, it has the notable feature 
that the third order (three generation) effects
such as CP violation can be calculated.

\section{Summary}

\hspace*{\parindent}
In this paper, we consider the method how to
approximates the neutrino oscillation probabilities 
in matter under three generations and
the obtained results are summarized as following.
\begin{enumerate}
\item 
In the framework of two generation neutrino
oscillation,
we discuss the applicable region of the perturbative
expansion on the small mixing angle in matter.
The result of the perturbation differs largely from
the exact numerical calculation in the MSW resonance point.
This means that non-perturbative effects are important
even for the neutrino oscillation in two generations.
\item 
We extend the method \cite{Takamura04, Kimura04}
to calculate the approximate formulas, in which 
non-perturbative  effects of the small parameters
$\Delta m_{21}^2/\Delta m_{31}^2$ and $\sin
\theta_{13}$, to all channels.
Under the conditions, $\theta_{23}=45^\circ$
and the symmetric matter density profile,
we derive simple approximate formulas of the
probabilities in all
channels by using the unitary relation.
Although all these approximate formulas are
expressed by the amplitudes
calculated within the framework of two generations,
it has a notable feature that the three generation
effects such as CP violation can also be calculated.
\item 
In the three generation neutrino oscillation with matter,
we investigate non-perturbative effects of the two
small parameters
$\Delta m_{21}^2/\Delta m_{31}^2$ and $\sin
\theta_{13}$.
We compare our approximate formulas with those
from the double expansion,
which include the terms up to the second order
in the low and high energy MSW resonance regions.
The obtained result is that the second order
formulas show large differences from the exact numerical
calculation, which
means that non-perturbative effects of the small
$\Delta m_{21}^2/\Delta m_{31}^2$ and
$\sin \theta_{13}$ become important in the MSW
resonance region.
\end{enumerate}
Finally, we describe two problems that we could
not fully address
in this paper, and which are tasks for future
research.
\begin{enumerate}
\item 
The approximate formulas in this paper are
derived by using the condition $\theta_{23}=45^\circ$, which
is the center value obtained from the atmospheric neutrino 
experiments.
However, but differences from this value may exist 
within 90\% confidence level.
\item 
The condition for the symmetric matter density
is satisfied in the 1-dimensional models, like the
PREM and the ak135f,
but the actual matter density, for example, that from
J-PARC to Beijing is not symmetric \cite{Shan03}.
Therefore, our aim for future work is, to derive
more sophisticated approximate formulas 
that hold not only in symmetric matter but 
in arbitrary matter as well.

\end{enumerate}
To solve the above two problems are the future works
This is now included in the upper sentence.

\section*{Acknowledgement}

We are grateful to H. Yokomakura, and T. Yoshikawa for
useful discussions and careful reading of our manuscript.
We would like to thank Prof. Wilfried Wunderlich
(Nagoya Inst. Technology) for helpful comments and
advice on English expressions.

\appendix

\section{General Feature of CP Dependence}

\hspace*{\parindent}
In this appendix we calculate the coefficients of
the probabilities in detail.
We show that the 2-3 mixing angle and the CP phase
are not affected by matter, from a different point 
of view as described in our previous paper
\cite{Yoko02}.
This result means that we only have to consider
the matter effects on four parameters $(\Delta
m_{21}^2, \theta_{12})$ and
$(\Delta m_{31}^2, \theta_{13})$.
By using this result, we can understand the matter
effects in three generations, which become complex 
compared with that in two generations.

\subsection{Remarkable Features of Effective Masses}

\hspace*{\parindent}
In this subsection, we show that $(\theta_{23},
\delta)$
do not affect the effective mass in three generation
Hamiltonian.
If we express the effective Hamiltonian in matter as
\begin{eqnarray}
H =
U{\rm diag}(0,\Delta_{21},\Delta_{31})U^{\dagger}
+{\rm diag}(a,0,0),
\end{eqnarray}
the equation of eigenvalue is given by
\begin{eqnarray}
{\rm det}(t-H)
&=& t^3 - (\Delta_{21} + \Delta_{31} + a)t^2
\nonumber \\
&+& (\Delta_{21}\Delta_{31}
+
a(\Delta_{21}(1-|U_{e2}|^2)+\Delta_{31}(1-|U_{e3}|^2)))t
- a\Delta_{21}\Delta_{31} |U_{e1}|^2 = 0,
\end{eqnarray}
and by solving this equation, we obtain the effective
masses as
\begin{eqnarray}
\lambda_1 & = & \frac{A}{3} -
\frac{1}{3}\sqrt{A^2-3B}S -
\frac{\sqrt{3}}{3}\sqrt{A^2-3B}\sqrt{1-S^2}\\
\lambda_2 & = & \frac{A}{3} -
\frac{1}{3}\sqrt{A^2-3B}S +
\frac{\sqrt{3}}{3}\sqrt{A^2-3B}\sqrt{1-S^2}\\
\lambda_3 & = & \frac{A}{3} +
\frac{2}{3}\sqrt{A^2-3B}S
\end{eqnarray}
\cite{Barger80}, where $A, B, C$ and $S$ are defined
by
\begin{eqnarray}
A & = & \Delta_{21} + \Delta_{31} + a \\
B & = & \Delta_{21}\Delta_{31}
+
a[\Delta_{21}(1-|U_{e2}|^2)+\Delta_{31}(1-|U_{e3}|^2)]
\\
C & = & a\Delta_{21}\Delta_{31} |U_{e1}|^2 \\
S & = & \cos\left[\frac{1}{3}\arccos\left(
\frac{2A^3-9AB+27C}{2\sqrt{(A^2-3B)^3}}\right)\right].
\end{eqnarray}
These effective masses depend only on the following
three
vacuum mixing angles
\begin{eqnarray}
|U_{e1}| = c_{12}c_{13}, \quad |U_{e2}| =
s_{12}c_{13},
\quad |U_{e3}| = s_{13}.
\end{eqnarray}
One can see from these equalities that the effective
masses are independent of
the 2-3 mixing angle $\theta_{23}$ and the CP phase
$\delta$.
Next, let us consider this result from a different
point of view.

\subsection{Decomposition of 2-3 mixing and CP Phase
from Hamiltonian}

\hspace*{\parindent}
In this section, we separate $\theta_{23}$ and
$\delta$ from the Hamiltonian and we study the
dependence of the
amplitudes on the two small parameters $\alpha$ and
$s_{13}$.
The Standard Parametrization is defined by
\begin{eqnarray}
U = O_{23}\Gamma O_{13}\Gamma^\dagger O_{12},
\end{eqnarray}
where the CP phase matrix $\Gamma$ is given by
\begin{equation}
\Gamma = {\rm diag}(1,1,e^{i\delta}).
\end{equation}
The CP phase matrix $\Gamma$ and the 1-2 mixing matrix
$O_{12}$ are commutable as
\begin{equation}
[\Gamma, O_{12}] = [\Gamma,{\rm
diag}(0,\Delta_{21},\Delta_{31})] = 0.
\end{equation}
Therefore, the Hamiltonian can be separated as
\begin{eqnarray}
H(t)
= U{\rm diag}(0,\Delta_{21},\Delta_{31})U^{\dagger}
+ {\rm diag}(a(t),0,0)
= O_{23}\Gamma H'(t)\Gamma^{\dagger}O_{23}^T
\label{decomposition-1},
\end{eqnarray}
where $H'(t)$ is defined by
\begin{equation}
H'(t) = O_{13}O_{12}{\rm
diag}(0,\Delta_{21},\Delta_{31})O_{12}^TO_{13}^T
+ {\rm diag}(a(t),0,0).
\end{equation}
This means that the 2-3 mixing and the CP phase can be
separated
from the part which includes the matter effects
$a(t)$.
\par
In the case of constant matter density profile, we
obtain
\begin{eqnarray}
{\rm det}(\lambda-H) = {\rm det}(\lambda-H'),
\end{eqnarray}
the 2-3 mixing angle and the CP phase do not affect
the eigenvalue
equation.
Accordingly, the effective masses are independent of
the 2-3 mixing
angle and the CP phase, which coincide with the result
obtained in the previous subsection.

\subsection{Exact CP and 2-3 mixing Dependence of
Oscillation Probabilities}

\hspace*{\parindent}
Here, let us consider the case in which we apply
the above discussion used in the Hamiltonian to the
amplitude.
Solving the Schrodinger eq. for the amplitude in
matter, we obtain
\begin{equation}
S(t) = {\rm T}\exp\left\{-i\int H(t) dt\right\}.
\end{equation}
By using this, we obtain
\begin{eqnarray}
S(t)
= {\rm T}\exp\left\{-i\int
O_{23}\Gamma H'(t)\Gamma^{\dagger}O_{23}^T dt\right\}
= O_{23}\Gamma {\rm T}\exp\left\{-i\int
H'(t) dt\right\} \Gamma^{\dagger}O_{23}^T
= O_{23} \Gamma S'(t)\Gamma^\dagger O_{23}^T
\end{eqnarray}
from (\ref{decomposition-1}).
Therefore, $S(t)$ satisfies
\begin{equation}
S(t) = O_{23} \Gamma S'(t)\Gamma^\dagger O_{23}^T.
\end{equation}
From this equation, we obtain
\begin{eqnarray}
P(\nu_e \to \nu_e) &=& C_{ee}, \\
P(\nu_\alpha \to \nu_\beta) &=&
A_{\alpha\beta} \cos\delta
+ B_{\alpha\beta} \sin\delta
+ C_{\alpha\beta},
\end{eqnarray}
when the initial or final state is $\nu_e$,
and
\begin{eqnarray}
P(\nu_\alpha \to \nu_\beta) &=&
A_{\alpha\beta} \cos\delta
+ B_{\alpha\beta} \sin\delta
+ C_{\alpha\beta}
+ D_{\alpha\beta} \cos2\delta
+ E_{\alpha\beta} \sin2\delta,
\end{eqnarray}
in the case of $\nu_\alpha, \nu_\beta=\nu_\mu,
\nu_\tau$ \cite{Yoko02}.
The final result is given by
\begin{eqnarray}
\hspace{-2em}
P(\nu_e \to \nu_e) \label{P_ee}
&=& C_{ee} = |S'_{ee}|^2, \label{C_ee} \\
\hspace{-1em}
P(\nu_e \to \nu_\mu) &=&
A_{e\mu} \cos\delta
+ B_{e\mu} \sin\delta
+ C_{e\mu}, \label{P_emu} \\
A_{e\mu}
&=& 2 {\rm Re}[{S}_{\mu e}^{'*} S'_{\tau e}] c_{23}
s_{23} ,
\hspace{0.5em} \label{A_emu} \\
B_{e\mu}
&=& - 2 {\rm Im}[{S}_{\mu e}^{'*} S'_{\tau e}] c_{23}
s_{23} ,
\hspace{0.5em} \label{B_emu} \\
C_{e\mu}
&=& |S'_{\mu e}|^2 c_{23}^2 + |S'_{\tau e}|^2
s_{23}^2,
\hspace{0.5em} \label{C_emu} \\
\hspace{-1em}
P(\nu_e \to \nu_\tau) &=&
A_{e\tau} \cos\delta
+ B_{e\tau} \sin\delta
+ C_{e\tau}, \label{P_etau}
\\
A_{e\tau}
&=& - 2 {\rm Re}[{S}_{\mu e}^{'*} S'_{\tau e}] c_{23}
s_{23},
\hspace{0.5em} \label{A_etau} \\
B_{e\tau}
&=& 2 {\rm Im}[{S}_{\mu e}^{'*} S'_{\tau e}] c_{23}
s_{23},
\hspace{0.5em} \label{B_etau} \\
C_{e\tau}
&=& |S'_{\mu e}|^2 s_{23}^2 + |S'_{\tau e}|^2
c_{23}^2,
\hspace{0.5em} \label{C_etau} \\
\hspace{-1em}
P(\nu_\mu \to \nu_\mu) &=&
A_{\mu\mu} \cos\delta
+ B_{\mu\mu} \sin\delta
+ C_{\mu\mu}
+ D_{\mu\mu} \cos2\delta
+ E_{\mu\mu} \sin2\delta \label{P_mm}
, \\
A_{\mu\mu}
&=& 2 {\rm Re}[({S}'_{\mu \mu} c_{23}^2 + S'_{\tau
\tau} s_{23}^2)^{*}
({S}'_{\tau \mu} + S'_{\mu \tau})] c_{23} s_{23} ,
\hspace{0.5em} \label{A_mm} \\
B_{\mu\mu}
&=& - 2 {\rm Im}[({S}'_{\mu \mu} c_{23}^2 + S'_{\tau
\tau} s_{23}^2)^{*}
({S}'_{\tau \mu} - S'_{\mu \tau})] c_{23} s_{23},
\hspace{0.5em} \label{B_mm} \\
C_{\mu\mu}
&=& |S'_{\mu \mu}|^2 c_{23}^4
+ |S'_{\tau \tau}|^2 s_{23}^4
+ (|S'_{\mu \tau}|^2 + |S'_{\tau \mu}|^2
+ 2{\rm Re}[{S}_{\mu \mu}^{'*} S'_{\tau \tau}])
c_{23}^2 s_{23}^2,
\hspace{0.5em} \label{C_mm} \\
D_{\mu\mu}
&=& 2{\rm Re}[{S}_{\tau \mu}^{'*}S'_{\mu \tau}]
c_{23}^2 s_{23}^2 ,
\hspace{0.5em} \label{D_mm} \\
E_{\mu\mu}
&=& 2{\rm Im}[{S}_{\tau \mu}^{'*}S'_{\mu \tau}]
c_{23}^2 s_{23}^2 ,
\hspace{0.5em} \label{E_mm} \\
\hspace{-1em}
P(\nu_\tau \to \nu_\tau) &=&
A_{\tau\tau} \cos\delta
+ B_{\tau\tau} \sin\delta
+ C_{\tau\tau}
+ D_{\tau\tau} \cos2\delta
+ E_{\tau\tau} \sin2\delta \label{P_tt}
, \\
A_{\tau\tau}
&=& - 2 {\rm Re}[({S}'_{\mu \mu} s_{23}^2 + S'_{\tau
\tau} c_{23}^2)^{*}
({S}'_{\tau \mu} + S'_{\mu \tau})] c_{23} s_{23} ,
\hspace{0.5em} \label{A_tt} \\
B_{\tau\tau}
&=& 2 {\rm Im}[({S}'_{\mu \mu} s_{23}^2 + S'_{\tau
\tau} c_{23}^2)^{*}
({S}'_{\tau \mu} - S'_{\mu \tau})] c_{23} s_{23} ,
\hspace{0.5em} \label{B_tt} \\
C_{\tau\tau}
&=& |S'_{\mu \mu}|^2 s_{23}^4
+ |S'_{\tau \tau}|^2 c_{23}^4
+ (|S'_{\mu \tau}|^2 + |S'_{\tau \mu}|^2
+ 2 {\rm Re}[{S}_{\mu \mu}^{'*} S'_{\tau \tau}])
c_{23}^2 s_{23}^2,
\hspace{0.5em} \label{C_tt} \\
D_{\tau\tau}
&=& 2 {\rm Re}[{S}_{\tau \mu}^{'*}S'_{\mu \tau}]
c_{23}^2 s_{23}^2 ,
\hspace{0.5em} \label{D_tt} \\
E_{\tau\tau}
&=& 2 {\rm Im}[{S}_{\tau \mu}^{'*}S'_{\mu \tau}]
c_{23}^2 s_{23}^2 ,
\hspace{0.5em} \label{E_tt} \\
\hspace{-1em}
P(\nu_\mu \to \nu_\tau) &=&
A_{\mu\tau} \cos\delta
+ B_{\mu\tau} \sin\delta
+ C_{\mu\tau}
+ D_{\mu\tau} \cos2\delta
+ E_{\mu\tau} \sin2\delta \label{P_mt}
, \\
A_{\mu\tau}
&=& - 2 {\rm Re}[({S}'_{\mu \mu} - S'_{\tau \tau})^{*}
({S}'_{\tau \mu} c_{23}^2 - S'_{\mu \tau}
s_{23}^2)]c_{23} s_{23} ,
\hspace{0.5em} \label{A_mt} \\
B_{\mu\tau}
&=& 2 {\rm Im}[({S}'_{\mu \mu} - S'_{\tau \tau})^{*}
({S}'_{\tau \mu} c_{23}^2 + S'_{\mu \tau} s_{23}^2)]
c_{23} s_{23} ,
\hspace{0.5em} \label{B_mt} \\
C_{\mu\tau}
&=& |S'_{\mu \tau}|^2 s_{23}^4 + |S'_{\tau \mu}|^2
c_{23}^4
+ (|S'_{\mu \mu}|^2 + |S'_{\tau \tau}|^2
- 2 {\rm Re}[{S}_{\mu \mu}^{'*} S'_{\tau
\tau}])c_{23}^2 s_{23}^2,
\hspace{0.5em} \label{C_mt} \\
D_{\mu\tau}
&=& - 2 {\rm Re}[{S}_{\tau \mu}^{'*}S'_{\mu \tau}]
c_{23}^2 s_{23}^2 ,
\hspace{0.5em} \label{D_mt} \\
E_{\mu\tau}
&=& - 2
{\rm Im}[{S}_{\tau \mu}^{'*}S'_{\mu \tau}] c_{23}^2
s_{23}^2.
\hspace{0.5em} \label{E_mt}
\end{eqnarray}

\end{document}